\documentclass[9pt,twocolumn,twoside]{pnas-new}
\usepackage[utf8]{inputenc}
\usepackage{natbib}

\templatetype{pnasinvited} 

\title{Revealing Robust Oil and Gas Company Macro-Strategies using Deep Multi-Agent Reinforcement Learning}

\author[a,1]{Dylan Radovic}
\author[a,b]{Lucas Kruitwagen} 
\author[c]{Christian Schroeder de Witt}
\author[a]{Ben Caldecott}
\author[d]{Shane Tomlinson}
\author[e]{Mark Workman}

\affil[a]{University of Oxford - Smith School of Enterprise and the Environment}
\affil[b]{University of Oxford - Institute for New Economic Thinking}
\affil[c]{University of Oxford - FLAIR, Department of Engineering Science}
\affil[d]{E3G}
\affil[e]{Imperial College London - Energy Futures Lab}

\leadauthor{Radovic} 

\authorcontributions{Please provide details of author contributions here.}
\authordeclaration{There are no competing interests here.}
\equalauthors{}
\correspondingauthor{\textsuperscript{1}To whom correspondence should be addressed. E-mail: dylan.radovic@gmail.com}

\keywords{ oil and gas majors $|$ international oil companies $|$ decision-making under deep uncertainty $|$ energy systems $|$ transition risks $|$ energy transition $|$sustainable finance $|$ energy investments $|$ climate scenario analysis $|$ robustness $|$ deep multi-agent reinforcement learning $|$ wargaming $|$ game theory $|$ general-sum games $|$ low-carbon transition $|$ investor engagement $|$ leveraged transition $|$ climate change

} 

\begin{abstract}
The energy transition potentially poses an existential risk for major international oil companies (IOCs) if they fail to adapt to low-carbon business models. Projections of energy futures, however, are met with diverging assumptions on its scale and pace, causing disagreement among IOC decision-makers and their stakeholders over what the business model of an incumbent fossil fuel company should be. In this work, we used deep multi-agent reinforcement learning to solve an energy systems wargame wherein players simulate IOC decision-making, including hydrocarbon and low-carbon investments decisions, dividend policies, and capital structure measures, through an uncertain energy transition to explore critical and non-linear governance questions, from leveraged transitions to reserve replacements. Adversarial play facilitated by state-of-the-art algorithms revealed decision-making strategies robust to energy transition uncertainty and against multiple IOCs. In all games, robust strategies emerged in the form of low-carbon business models as a result of early transition-oriented movement. IOCs adopting such strategies outperformed business-as-usual and delayed transition strategies regardless of hydrocarbon demand projections. In addition to maximizing value, these strategies benefit greater society by contributing substantial amounts of capital necessary to accelerate the global low-carbon energy transition. Our findings point towards the need for lenders and investors to effectively mobilize transition-oriented finance and engage with IOCs to ensure responsible reallocation of capital towards low-carbon business models that would enable the emergence of fossil fuel incumbents as future low-carbon leaders. 
\end{abstract}

\dates{This manuscript was compiled on \today~based on sources published at SSRN (\url{https://papers.ssrn.com/sol3/papers.cfm?abstract_id=3933996}) on Sept 30, 2021.}

\doi{}
\begin{document}

\maketitle
\thispagestyle{firststyle}
\ifthenelse{\boolean{shortarticle}}{\ifthenelse{\boolean{singlecolumn}}{\abscontentformatted}{\abscontent}}{}

\noindent The strategies adopted by major oil and gas companies will help to determine their contribution to the transformation of the global energy system. Since the 2014 oil price collapse, declining performances of international oil companies (IOCs) have highlighted issues faced by their business model~\citep{stevens_international_2016} as evident with increasing upstream capital costs~\citep{ihs_markit_energy_2020,carbon_tracker_sense_2016} and declining returns~\citep{fitz_winning_2019,novel_investor_annual_2020}, dividend inflation~\citep{hipple_ieefa_2020} and waves of asset write-downs~\citep{eaton_2020_2020}. The 9\% drop in oil demand due to the COVID-19 pandemic has amplified these issues with significant losses~\citep{hiller_pandemic_2021}, prompting cuts to capital expenditures~\citep{oilgas_global_2020} and further valuation declines~\citep{ambrose_seven_2020}. Furthermore, efforts to achieve the goals set out by the Paris Agreement~\citep{unfccc_paris_2018} to combat climate change are gaining momentum as economies look to accelerate the low-carbon energy transition~\citep{larsen_2020_2021,unfccc_commitments_2020}. 

The potential magnitude of energy transition risks~\citep{tcfd_task_2016} (e.g. demand shocks, impairments) are concerning investors~\citep{climate_action_100_climate_2020} of fossil fuel companies. Coupled with IOCs’ potential capital misallocation~\citep{landell-mills_are_2018}, these risks could strand oil and gas assets~\citep{carbon_tracker_mind_2018,carbon_tracker_wasted_2013,mcglade_geographical_2015}, reduce industry revenues by potentially trillions of dollars~\citep{lewis_stranded_2014}, and create market instability as a result of reduced asset valuations~\citep{carney_open_2019,spedding_oil_2013,pri_preparing_2020,fattouh_energy_2019}. 
In response, peer-reviewed literature~\citep{stevens_international_2016,west_energy_2019,boon_climate_2019,pickl_renewable_2019,zhong_contours_2018,stevens_geopolitical_2019,eccles_implementing_2019} and industry reports~\citep{iea_oil_2020,johnston_role_2020,asmelash_international_2021,beck_how_2021} regarding oil and gas companies in the energy transition echo the same sentiment: the risks inherent to an energy transition could lead to the collapse of IOC business models, particularly the Majors\footnote{In this work, we define the oil and Majors as ExxonMobil, Chevron, BP, Total, Shell, and Eni.}, if they fail to adapt.

To mitigate downside risks, studies suggest several potential low-carbon business model opportunities and strategies as pathways towards transition- and climate-compatibility. Achieving a successful low-carbon transition requires IOCs to execute a challenging balancing act—that is, generate the necessary short-term returns for shareholders while investing in low-carbon businesses for future profitability. Although the Majors have recently announced pathways to cut carbon emissions and increase transition-oriented spending~\citep{bp_bp_2020,totalenergies_total_2020,shell_shell_2021,eni_reducing_2021}, criticisms have emerged due to likely incompatibility with the climate goals and insufficient low-carbon capital expenditures~\citep{carbon_tracker_absolute_2021}. The studies and criticisms~\citep{iea_oil_2020,asmelash_international_2021,carbon_tracker_absolute_2020,shojaeddini_oil_2019,fletcher_beyond_2018,carbon_tracker_2_2017,carbon_tracker_fault_2020,carbon_tracker_balancing_2019,carbon_tracker_breaking_2019,iea_stepup_2020,dnv_turmoil_2021} contributing to a widespread narrative that the Majors must change are, however, predicated on a range of energy futures scenarios with low hydrocarbon demand projections and widely varying and often not very transparent assumptions. 
Rigorous assessments of these energy futures and the robustness of their conclusions and what these mean for the future of the Majors remains lacking. As a result, analysis of risks and rewards of changing business models at different times and under a range of market conditions are largely missing from the literature and its low-carbon consensus. This paper seeks to close this gap.

Here, we develop a data-driven approach to reveal and assess emergent IOC strategies robust\footnote{In this work, robust IOC strategies are defined as strategies that minimize downside risks that may arise from market uncertainty and competitor counter-strategies. A robust strategy is not necessarily one that leads to the greatest gains; multiple robust strategies may be present in a single game. Hence, the framing of 2DP as a general-sum game to allow for ``win-win'' scenarios (see Methods). The main IOCs training using the league mechanism described in Methods will boast robust strategies if they successfully mitigate downside risks, as indicated by the applied reward function (see Appendix Table A.\ref{tbl:ioc_reward_function}). Exploiter IOCs, on the other hand, do not yield robust strategies as they seek to optimize for opponent weaknesses, not energy futures and competitor uncertainty.} to market and competitor uncertainty. To achieve this, we built a multi-agent system that solves an oil and gas majors wargame across the most recent collection of integrated assessment model (IAM) scenarios with deep reinforcement learning (Figure \ref{fig:wargame}, Appendix Table A.\ref{tbl:global_scenario_metrics}). Agents, acting as IOCs in market competition, were trained to compute an approximate best-response to varying market conditions and exploitative strategies along a $30$-year time horizon. 
This work builds upon several studies regarding oil and gas companies in the context of climate-related risks and the energy transition as well as the utilization of multi-agent learning and deep reinforcement learning to explore emergent, robust agent behavior. 

Early climate-related risk work explored the impact carbon budgets will have on fossil fuel companies~\citep{carbon_tracker_wasted_2013,carbon_tracker_unburnable_2011,meinshausen_greenhouse-gas_2009}. Institutions echoed these sentiments, calling for a massive reallocation of capital towards low-carbon solutions~\citep{carney_breaking_2015} and the disclosure of climate-related risks, physical and transitional, most pertinent to business activity~\citep{tcfd_task_2016}. Simultaneously, energy pathways and scenarios using IAMs have been proposed, and are continually updated, to guide decision-makers on decarbonization strategies~\citep{huppmann_iamc_2018,mai_re-assume_2013,pye_energy_2017,iea_net_2021}.  Of significant importance to this work, studies exploring and quantifying transition risks with respect to the oil and gas industry that arise from these pathways have been elusive. This is largely due to the limitations of present scenario-analysis~\citep{paltsev_energy_2017} as well as the policy insight shortcomings in the contexts of uncertainty~\citep{workman_decision_2020}. Recent studies have provided analyses on the state of oil and gas companies in the energy transition as well as suggested potential strategic responses~\citep{iea_net_2021,asmelash_international_2021,beck_how_2021,dalman_carbon_2020,carbon_tracker_2_2017_2,bloombergnef_integrated_2021}. Tangible upside and downside risks of their recommended strategies, however, are largely missing due to the studies’ linear assumptions and focuses on a singular energy future.

The $2$ Degrees Pathways (2DP) wargaming tool~\citep{tomlinson_crude_2018} sought to fill this research gap and inform stakeholder thinking around the macro-strategies oil and gas companies can take to become climate-compatible by simulating oil and gas companies in competition. Oil and gas competitive game theory simulations are used to enhance company strategic decision-making~\citep{koch_game_2011,castillo_decision-making_2013,chang_oil_2014,schitka_applying_2014,azmi_application_2017,willigers_game_2009}. Applying these conventional methods to discover effective company strategies, however, proves intractable due to the 2DP’s complexity as a high-dimensional continuous control problem (see Methods). 

Advances in reinforcement learning have overcome game-theoretic challenges, successfully training agents to achieve superhuman-level performance in complex games such as Backgammon~\citep{tesauro_temporal_1995} and Go~\citep{silver_mastering_2016}, StarCraft~\citep{vinyals_grandmaster_2019} and Dota~\citep{openai_dota_2019}. Of particular importance to this work, AlphaStar (StarCraft) and OpenAI Five (Dota) demonstrated that the combination of deep reinforcement learning and multi-agent learning can prove powerful in generating complex, robust agent behavior within high-dimensional continuous control environments. To the best of our knowledge, there is not yet a deep multi-agent reinforcement learning model that solves a wargaming tool relevant to the oil and gas industry in the energy transition until this work.

\section*{Solving a Wargame.}

\begin{figure*}
\centering
\includegraphics[width=.9\linewidth]{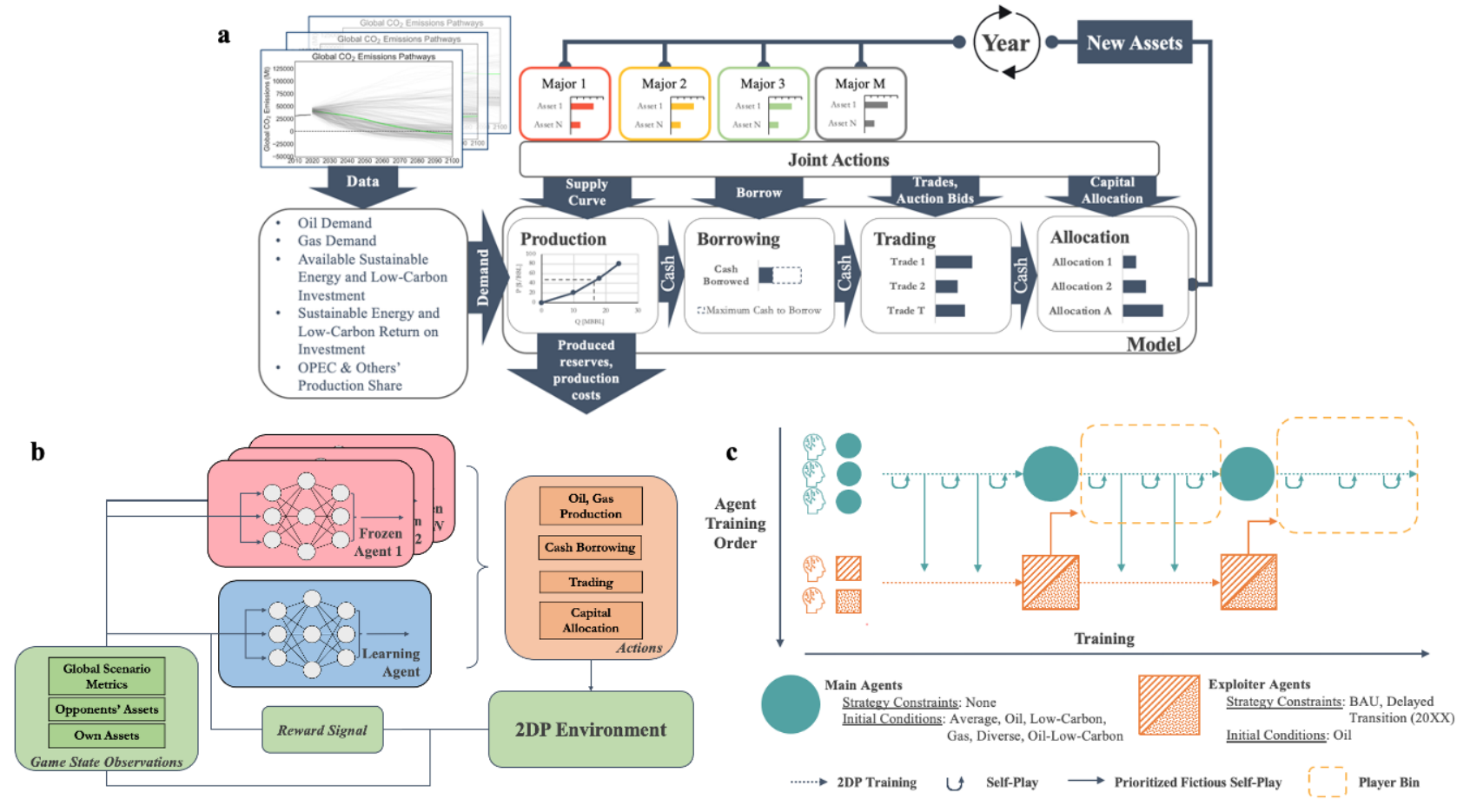}
\caption{The model and training setup. \textbf{a,} The 2DP wargame involves multiple players competing as IOCs, resemblant of major oil and gas companies, within varying energy futures scenarios. IOCs select actions based on the global scenario metrics given to them, as well as information about themselves and their opponents, that dictate a year’s production, cash borrowing, trading and low-carbon asset auction bidding, and capital allocation (Appendix Tables A.\ref{tbl:energy_scenarios}). The game advances to the next year after the allocation stage. Upon reaching the endgame, a new energy scenario is chosen to alter dynamics of the next game. \textbf{b,} We train our 2DP IOCs with deep reinforcement learning. The learning IOC collects game experience in the 2DP wargame via game state observations and a reward signal to update its strategy. Frozen (non-learning) IOCs are chosen by the multi-agent learning mechanism at play and compete against the learning IOC. \textbf{c,} We use a league training system involving two learning IOCs to address game-theoretic challenges. Main IOCs do not have strategy constraints but vary in initial asset distributions and are initially trained against copies of themselves. Exploiter are added as opponents in subsequent iterations. Exploiter IOCs, constrained to certain strategies (e.g. BAU, delayed transition $2030$).}
\label{fig:wargame}
\end{figure*}

Core to this work, we use a variant of the 2DP wargaming tool61 game environment to simulate IOCs in competition within varying energy futures scenarios (Figure \ref{fig:wargame}a). Solving 2DP to discover novel behavior, however, posits game-theoretic challenges due to the game’s non-linear payoffs, continuous high dimensional state and action space, planning horizon, and emergent counter-strategies. Furthermore, 2DP participants must balance economic decisions simultaneously given only incomplete and imperfect information of the game environment and rival players. To address these challenges and complexities, our model applies a combination of deep reinforcement learning and multi-agent learning. Details on the game environment as well as learning techniques employed are elaborated in Methods.

In the 2DP wargame, IOCs respond to varying energy futures scenario metrics, such as oil demand and low-carbon return on investment (ROI), by developing individual robust macro-strategies. These strategies emerge in the form of a combination of selected available actions: choose hydrocarbon production levels; invest in oil and gas exploration; invest in oil and gas development; invest in sustainable energy and low-carbon technologies or business models; allocate dividends; and access credit to adjust capital structure (Appendix Tables A.\ref{tbl:global_scenario_metrics},\ref{tbl:ioc_game_state_observations},\ref{tbl:ioc_action_space1}, and \ref{tbl:ioc_action_space2}). An IOC’s actions, as well as those of others, dictate the game’s core dynamics (Figure \ref{fig:wargame}a). ‘Winning’ the game requires an IOC to maximize shareholder value over its opponents via dividends payouts. 2DP serves as a suitable testbench to address our research questions as the creators balanced the game’s cogency and verisimilitude to enable a range of behaviors across potential energy futures and macro-strategies.

We use reinforcement learning, the computational approach to learning from environment interaction~\citep{sutton_reinforcement_2018}, to enable a 2DP IOC to discover scenario-robust strategies that maximize unlevered dividend payouts (see Methods). The learning IOC achieves this by computing approximate best-responses to limited game state information, to best mimic real-life market competition, and a real-valued reward signal—a value predicated on the efficacy of its developed strategy towards achieving its goal (see Appendix Table A.\ref{tbl:ioc_reward_function}). Our training method follows an independent learning (InRL) approach whereby a single learning IOC competes against non-learning IOCs (i.e. IOCs playing previously learned strategies) that are chosen by the multi-agent learning mechanism described below (Figure \ref{fig:wargame}b). We equip our IOCs with deep neural networks to alleviate concerns regarding 2DP’s non-linearities, high dimensional state and action spaces, and planning horizons.

Discovering novel, robust strategies in 2DP is difficult due to its complex dynamics and game-theoretic challenges. Self-play reinforcement learning algorithms train a learning agent by simulating play against itself. Successful applications of self-play to achieve superhuman-level performance are seen in games such as Backgammon~\citep{tesauro_temporal_1995} and Go~\citep{silver_mastering_2016}, StarCraft~\citep{vinyals_grandmaster_2019} and Dota~\citep{openai_dota_2019}. Despite resulting in emergent, complex behavior, agents trained with self-play are susceptible to forgetting~\citep{balduzzi_open-ended_2019} (i.e. improve against itself but fail to win against past versions) and training imbalances~\citep{hernandez_generalized_2019}. Fictitious self-play~\citep[FSP]{heinrich_fictitious_2015} solves these issues by uniformly sampling opponents from past versions of the learning agent. DeepMind~\cite{vinyals_grandmaster_2019} extended this approach with prioritized fictitious self-play (PFSP) to train against a non-uniform mixture of opponents by focusing on the most difficult of agents. To address game-theoretic challenges and make our IOCs’ strategies more robust, we use a combination of the aforementioned self-play algorithms to create a league of learning IOCs, similarly used by DeepMind to discover novel strategies in a real-time strategy game~\citep{vinyals_grandmaster_2019}. 

Our 2DP-league mechanism involves two distinct learning IOCs that differ in the distribution of opponents they play against, when training parameters are reset, and the availability of actions (Figure \ref{fig:wargame}c). Our main IOCs train as energy companies with the goal of learning robust, investor-attractive strategies utilizing a mixture of self-play and PFSP. Exploiter IOCs, on the other hand, train as energy companies with the goal of finding strategy weaknesses in the latest versions of the main IOCs. Unlike the main IOCs, constraints are imposed on the actions available to them so they become dominant with respect to the strategies they are forced to play. We focus on two constrained strategies: BAU and delayed transition. The former restricts an IOC from entering low-carbon markets thereby optimizing strategies in oil and gas markets, the latter does the same until a given year (e.g. $2030$). This method encourages main IOCs, later trained against the exploiters, to adapt to such extreme strategies. Unlike the exploiter IOCs, training parameters for the main IOCs never reset, enabling continuous learning.

We train eight IOCs—six main, two exploiters (one BAU, one $2030$ delayed transition)—in a six player 2DP game by training each IOC consecutively against five non-learning opponents as described by the multi-agent mechanism (see Methods). IOCs are initiated with the average 2DP-relevant balance sheet items drawn from six oil Majors’ annual reports~\citep{sp_sp_2020} to best represent market conditions (Appendix Table A.\ref{tbl:ioc_balance_sheet_distribution}). The six main IOCs begin with differing asset distributions to demonstrate how an initial market dominance, or diversity, may affect strength of overall strategies. Our exploiter IOCs each begin with greater oil market share, optimizing either a BAU or delayed transition strategy. An IOC’s training iteration plays through $408$ unique energy futures scenarios drawn from the Integrated Assessment Model Community (IAMC) and International Institute for Applied Systems Analysis (IIASA) $1.5^\circ C$ explorer database~\citep{huppmann_iamc_2018} and the IEA's Net-Zero report~\citep{iea_net_2021} (Appendix Table A.\ref{tbl:energy_scenarios}). The IAMC/IIASA collection was chosen as it is the most recently updated ensemble of scenarios to span a range of energy futures within varying degree-warming scenarios, from below $1.5^\circ C$ to above $4^\circ C$. Each main IOC is trained for five iterations. 

\section*{Emergent, robust IOC strategies.}

\begin{figure*}
\centering
\includegraphics[width=.9\linewidth]{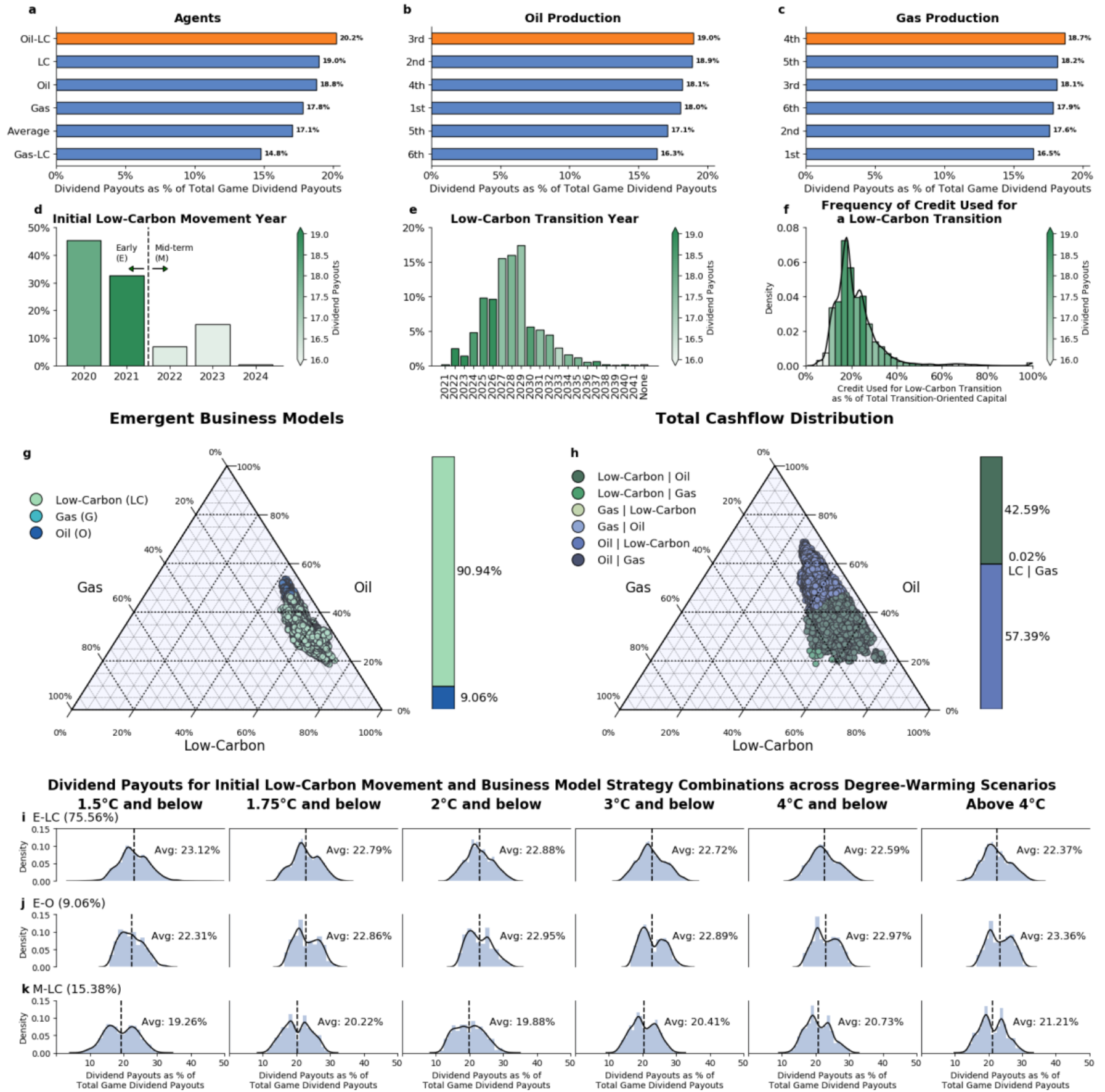}
\caption{ \textbf{Emergent strategies from main IOCs.} \textbf{a-c}, Resultant dividend payouts as compared to total dividend
payouts within each game for \textbf{(a)} each main IOC by initial asset portfolio, \textbf{(b)} top oil producers, \textbf{(c)} top gas producers. \textbf{d-f}, Frequency of low-carbon strategy emergence, and their average resultant dividend payouts as compared to total dividend payouts within each game, with respect to \textbf{(d)} the first year in which an IOC's low-carbon capital expenditures eclipse hydrocarbon capital expenditure, \textbf{(e)} the year an IOC's low-carbon assets are valued higher than their hydrocarbon reserves, \textbf{(f)} the significance of credit used to facilitate a complete low-carbon transition as shown in \textbf{(e)}. \textbf{g-h}, Emergent main IOC strategies with respect to \textbf{(g)} business model as determined by which market an IOC allocates capital to most, \textbf{(h)} the total cashflow distribution whereby the preceding market indicates the dominant source of income (e.g. Oil | Low-Carbon indicates oil cashflows are greater than low-carbon cashflows). \textbf{i-j}, Distribution of dividend payouts as compared to total dividend payouts within a given game separated by six degree-warming scenarios ($1.5^\circ C \pm 4^\circ C$) for all emergent strategies demonstrating \textbf{(i)} low-carbon business models as a result of early low-carbon movement (E-LC), \textbf{(j)} oil-focused business models with early movement towards low-carbon (E-O), \textbf{(k)} low-carbon business models as a result of mid-term low-carbon movement (M-LC). Such strategy combinations were obtained by drawing insight from \textbf{(d)} and \textbf{(e)}. Each E-LC and M-LC strategy combination boasted positive rewards signals, indicating such strategies successfully mitigated debt engulfment risks (see Appendix Table A.\ref{tbl:ioc_reward_function}).}
\label{fig:emergent_strategies}
\end{figure*}

After training, we evaluated the emergent strategies of the main IOCs when competing amongst varying combinations of themselves and the two exploiter IOCs, sampled from the league. Doing so required testing across $11,424$ games ($28$ unique opponent match-ups, $408$ energy scenarios), with each IOC playing $8,568$ games ($21$ unique opponent match-ups, $408$ energy scenarios). Matches were played under similar conditions as seen in training yet with less scenario metric noise (Appendix Tables A.\ref{tbl:global_scenario_metrics}, and \ref{tbl:ioc_game_state_observations}). IOCs played mixed strategies as non-deterministic play best represents real-life market competition (see Methods). To assess performances between strategies, we compared their individual cumulative unlevered dividend payouts with the total amount of unlevered dividend payouts allocated by each IOC across all games.

We observe whether greater initial market share, or diversity, will result in higher total dividend payouts (Figure \ref{fig:emergent_strategies}a). IOCs initialized with greater amounts of low-carbon assets, apart from Gas-LC, tend to perform better than those with initial portfolios focused on hydrocarbon markets. While the Oil-LC IOCs boast strategies that best its competitors, the greater market positioning in low-carbon markets proves more advantageous than equivalent positioning in oil as evidenced by the LC IOC’s performance over its Oil IOC peer. Moreover, we find that IOCs initialized with competitive advantages in gas markets noticeably underperform IOCs with advantages focused on the low-carbon or oil markets

To evaluate the IOCs’ tactics across the oil, gas, and low-carbon markets, we examined several, distinct categories with mutually exclusive strategies (Figure \ref{fig:emergent_strategies}b-f). With respect to oil markets, IOCs cutting production to quickly exiting the market risk underperformance. Increasing gas production, however, does not prove a favorable strategy for the IOCs, suggesting the carbon-intensive asset’s low returns. Observing low-carbon market behavior, we find emergent IOC strategies demonstrate considerable movement towards low-carbon business models by $2024$ with greater preferences towards allocating most capital towards low-carbon within the first two years of play ($76\%$, Figure \ref{fig:emergent_strategies}d). From a dividends perspective, IOCs that move into low-carbon early\footnote{We define ‘early’ (E) movers as IOCs that allocate most capital expenditures toward low-carbon within the first three years of play; ‘mid-term’ (M) movers are IOCs that exhibit this behavior after $2022$, but before $2025$.} outperform mid-term movers. Considerable capital movement towards low-carbon continues as $99.99\%$ of IOC strategies focus on completely transitioning to low-carbon business models—that is, when an IOC’s low-carbon assets are valued higher than their hydrocarbon reserves—a majority of which occurring within the mid- to late-$2020$s ($74\%$, Figure \ref{fig:emergent_strategies}e). Moreover, we discover that the use of credit, coupled with hydrocarbon production returns, was instrumental in facilitating a complete, robust low-carbon transition as well as developing high-dividend payout strategies (Figure \ref{fig:emergent_strategies}f). 

The aforementioned energy market tactics resulted in substantial similarities across all strategies’ endgame business models and total cash flow distribution (Figure \ref{fig:emergent_strategies}g,h). Observing capital allocation behavior across the three markets, we find that $90.94\%$ of emergent IOC strategies allocate considerable transition-oriented capital to facilitate transitions towards low-carbon business models by the endgame (Figure \ref{fig:emergent_strategies}g). Oil-focused business models accounted for the remaining $9.06\%$. To fund these energy investments, the IOCs’ main source of income are split between oil and low-carbon cashflows (Figure \ref{fig:emergent_strategies}h). For $57.39\%$ of emergent strategies, oil production generated a majority of an IOC’s income with supporting revenues from the acquired low-carbon assets. Remaining strategies ($42.61\%$) resulted in incomes with greater low-carbon revenues, outweighing both oil and gas cashflows. 

We further examine the performance of emergent strategy combinations comprising of initial low-carbon movement (Figure \ref{fig:emergent_strategies}d) and business model (Figure \ref{fig:emergent_strategies}g) strategies for all IOCs across several degree-warming scenarios (Figure \ref{fig:emergent_strategies}i-k). With $51,408$ opportunities ($6$ main IOCs, $21$ matchups, $408$ energy scenarios) to demonstrate trained, robust behavior, three distinct strategy combinations emerged: early and mid-term movement towards low-carbon both resulting in a low-carbon business model (E-LC, M-LC, respectively) and early movement towards low-carbon yet with total capital focused on sustaining oil business models (E-O). E-LC, M-LC, and E-O strategy combinations emerged across $75.56\%$, $15.38\%$, and $9.06\%$ of the $51,408$ opportunities. On average, E-O strategy combinations tend to outperform their E-LC and M-LC peers across most degree-warming scenarios and variations of competing strategy combinations, including the exploiters. In $1.5^\circ C$ and below degree-warming scenarios, however, E-LC IOCs tend to allocate the greatest share of dividends pointing towards the strategies' compatibility with climate policies. The performances of the E-O and M-LC IOCs increase with degree-warming scenarios, owing to their continued hydrocarbon production focuses that present greater upside risks with increasing demand curves.

\begin{figure*}
\centering
\includegraphics[width=.9\linewidth]{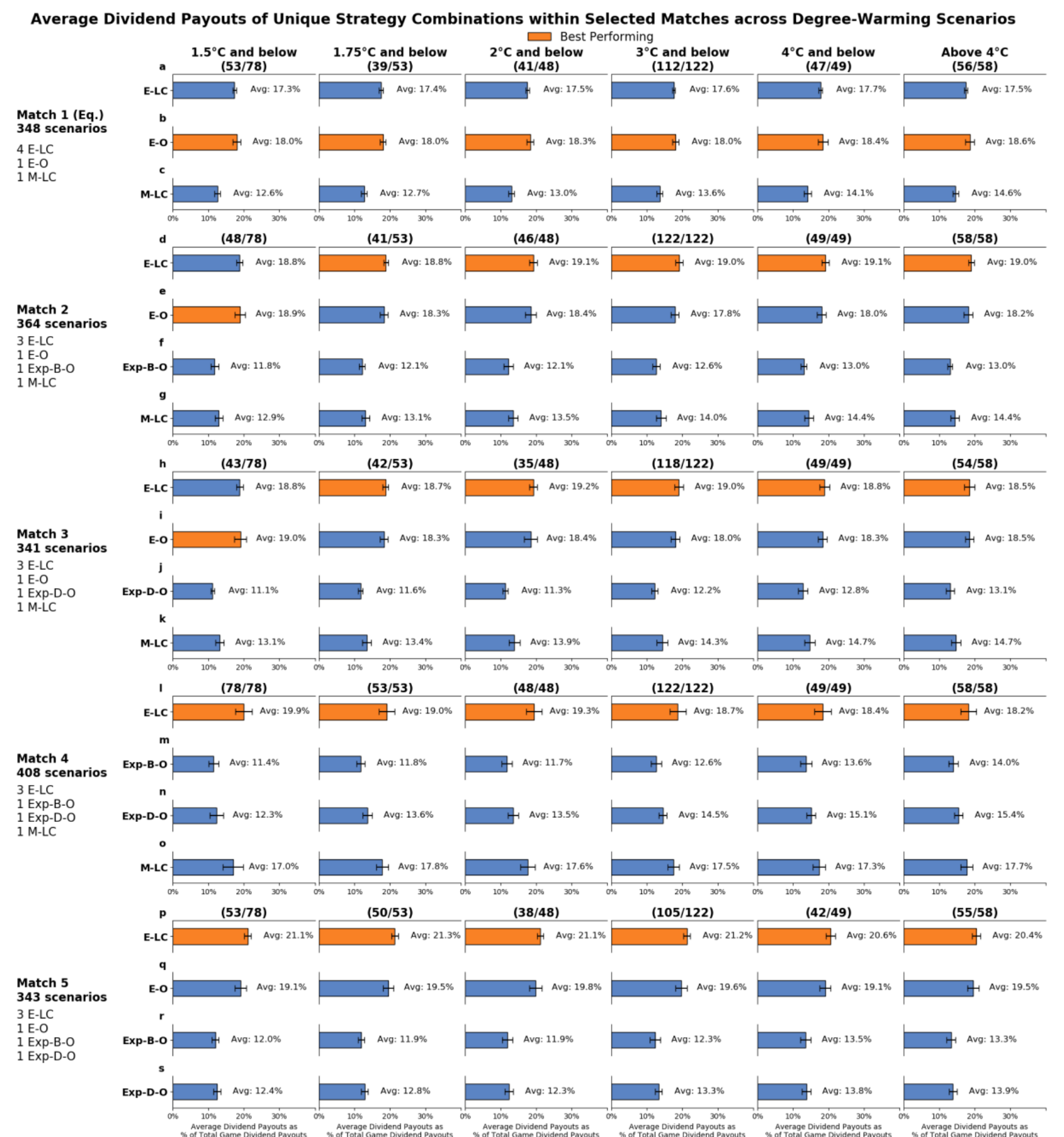}
\caption{ \textbf{Average dividend payouts of initial low-carbon movement (Figure \ref{fig:emergent_strategies}d) and business model (Figure
\ref{fig:emergent_strategies}g) strategy combinations within selected matches of similar or differing strategy combinations across degreewarming scenarios.} \textbf{a-c}, Average dividend payouts as compared to total dividend payouts seen in Match 1 (main
IOC approximate mixed-strategy Nash equilibrium) for the average \textbf{(a)} E-LC, \textbf{(b)} E-O, \textbf{(c)} M-LC strategy
combinations. \textbf{d-g}, Average dividend payouts as compared to total dividend payouts seen in Match 2 for the average
\textbf{(d)} E-LC, \textbf{(e)} E-O, \textbf{(f)} Exploiter-B-O, \textbf{(g)} M-LC strategy combinations. \textbf{h-k}, Average dividend payouts as compared
to total dividend payouts seen in Match 3 for the average \textbf{(h)} E-LC, \textbf{(i)} E-O, \textbf{(j)} Exploiter-D-O, \textbf{(k)} M-LC strategy
combinations. \textbf{l-o}, Average dividend payouts as compared to total dividend payouts seen in Match 4 for the average
\textbf{(l)} E-LC, \textbf{(m)} Exploiter-B-O, \textbf{(n)} Exploiter-D-O, \textbf{(o)} M-LC strategy combinations. \textbf{p-s}, Average dividend payouts as
compared to total dividend payouts seen in Match 4 for the average \textbf{(p)} E-LC, \textbf{(q)} E-O, \textbf{(r)} Exploiter-B-O, \textbf{(s)}
Exploiter-D-LC. Most matches did not occur across all $408$ scenarios due to sensitivity of strategies in response to
different game scenario metrics (e.g. a single M-LC strategy combination seen in Match 1 initiated earlier lowcarbon movement in certain scenarios thereby becoming an E-LC strategy combination). See Appendix Figure B.\ref{fig:average_dividend_payouts_appendix}
for additional unique matches, with greater than $300$ unique scenarios played, and their strategy combinations'
resulting dividend payout performance.}
\label{fig:average_dividend_payout}
\end{figure*}

We further examine the combination of low-carbon movement timing and business
model strategies as they perform against varying opponent combinations of themselves, and
exploiters, across different degree-warming scenarios (Figure \ref{fig:average_dividend_payout}). Exiting a market (e.g. oil, gas) to enter development into a new one (e.g. low-carbon) is a possible disadvantage for early movers as it creates exploitable gaps in the former market~\citep{lieberman_first-mover_1988}. This disadvantage is of particular
concern to IOCs in the event of high, continuously increasing hydrocarbon demand. The
equilibrium match, containing only the main IOCs, supports this notion as the average E-O IOCs
outperform the average E-LC and M-LC IOCs in most degree-warming scenarios as IOCs
focusing on oil development benefit from hydrocarbon market gaps (Figure \ref{fig:average_dividend_payout}a-c). Beyond
equilibrium and involving exploitive strategies, however, E-LC IOCs outperform most variations
of E-O, M-LC, Exp-B-O (BAU IOC), and Exp-D-O (delayed transition IOC) strategy
combination matchups and degree-warming scenarios with (Figure \ref{fig:average_dividend_payout}d-s, Appendix Figure B.\ref{fig:average_dividend_payouts_appendix}).
The E-LC strategy's robustness is most apparent in climate-compatible scenarios ($1.5\circ C$ and
below, $1.75^\circ C$ and below, $2^\circ C$ and below), apart from Match $2$ and $3$'s $1.5^\circ C$ scenarios, whereby
its dividend allocations remain higher than that of the next-best performer by margins ranging
from $1.6-10.4$\%. Notably, performances of exploiter IOCs increase with levels of degreewarming yet fail to payout higher dividends than any of their peers. While E-LC IOCs suffer
from underperformance in equilibrium, the early low-carbon movers effectively mitigate
concerns of exploitation from Exp-B-O and Exp-D-O strategy combinations.

\begin{figure*}
\centering
\includegraphics[width=.9\linewidth]{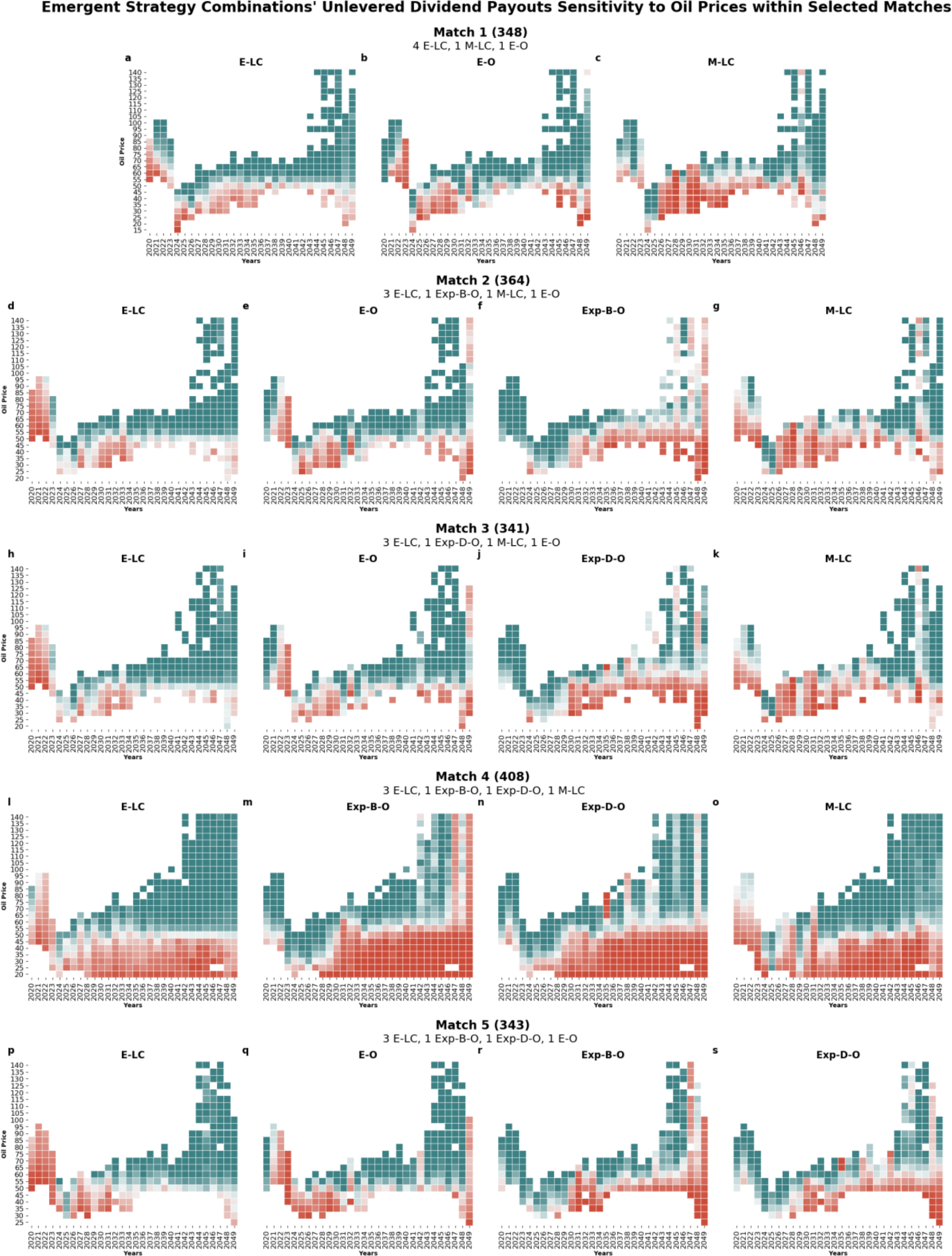}
\caption{\textbf{The sensitivity of yearly dividend payouts to varying oil prices for strategy combinations within five
unique matches.} \textbf{a-c}, Sensitivity of yearly dividend payouts to varying oil prices seen in Match 1 for average \textbf{(a)} ELC, \textbf{(b)} E-O, \textbf{(c)} M-LC strategy combinations. \textbf{d-g}, Sensitivity of yearly dividend payouts to varying oil prices seen in Match $2$ for average \textbf{(d)} E-LC, \textbf{(e)} E-O, \textbf{(f)} Exploiter-B-O, \textbf{(g)} M-LC strategy combinations. \textbf{h-k}, Sensitivity of
yearly dividend payouts to varying oil prices seen in Match $3$ for average \textbf{(h)} E-LC, \textbf{(i)} E-O, \textbf{(j)} Exploiter-D-O, \textbf{(k)}
M-LC strategy combinations. \textbf{l-o}, Sensitivity of yearly dividend payouts to varying oil prices seen in Match $4$ for
average \textbf{(l)} E-LC, \textbf{(m)} Exploiter-B-O, \textbf{(n)} Exploiter-D-O, \textbf{(o)} M-LC strategy combinations. \textbf{p-s}, Sensitivity of yearly
dividend payouts to varying oil prices seen in Match $5$ for average \textbf{(p)} E-LC, \textbf{(q)} E-O, \textbf{(r)} Exploiter-B-O, \textbf{(s)} M-LC.
See Appendix Figure B.2 \ref{fig:emergent_strategies_combinations_appendix} for additional unique matches.}
\label{fig:emergent_strategy_combinations}
\end{figure*}

Oil price forecasts are critical for internal stress testing of hydrocarbon business models
to oil price volatility~\citep{carbon_tracker_sense_2016} and are typically averaged to a single price for several years~\citep{carbon_tracker_fault_2020}. We
compare the upside and downside risks of the emergent strategy combinations within selected
matches, in parallel with Figure 3\ref{fig:average_dividend_payout}, when exposed to varying oil prices. From a high-level, in equilibrium, E-LC IOCs are less exposed to the downside risks of oil price drops, yet are unable
to attain the potential gains of higher oil prices similar to E-O IOCs (Figure \ref{fig:emergent_strategy_combinations}a,b). The addition of exploiter IOCs, however, challenges the E-O IOC's strategies resulting in decreased
possibility of gains with high oil prices and increased susceptibility to low oil prices (Figure
\ref{fig:emergent_strategy_combinations}e,i,q). E-LC IOCs successfully mitigate downside risks as well as maintain, and, in some cases, even increase, upside risks when compared to the E-O IOCs (Figure \ref{fig:emergent_strategy_combinations}d,h,p). This is largely due to the oil-focused IOCs' inabilities to benefit from market gaps when exploiter IOCs are involved.

Examining the sensitivity of the exploiter IOC's dividend payouts to oil prices, we find 
Exp-B-O strategies can maximize their upside risks, until the late $2020$s, yet fail to mitigate downside risk exposure to low oil prices (Figure \ref{fig:emergent_strategy_combinations}f,m,r). Moreover, Exp-B-O strategies' dividend payouts fall sharply towards the endgame despite high oil prices as compared to its adversaries. The sudden fallout points towards the Exp-B-O IOC's inability to replenish reserves and the main IOC's buildup of high low-carbon returns as a result of earlier transition strategies. Exp-D-O IOCs are unable to balance the upside and downside risks as delaying movement into low-carbon risks failure to facilitate a transition and acquire desirable returns in a timely manner.

Furthermore, we investigate the dividend dynamics of the main IOCs, particularly in the
early game. E-LC and E-O IOCs noticeably cut dividend payouts within the first four years of
play, regardless of the selected matchup. M-LC IOCs, on the other hand, delay dividend cuts as
the companies focus on business-as-usual strategies before reallocating capital towards lowcarbon in later years. The early dividend cuts by the early low-carbon movers highlight the need
too quickly reallocate capital towards building low-carbon economies that provide meaningful
mid- and long-term returns. In the case of E-LC IOCs, cutting dividends over consecutive years
in the early-game to elevate long-term returns risks displeasing investors in the near-term.

\begin{figure*}
\centering
\includegraphics[width=.9\linewidth]{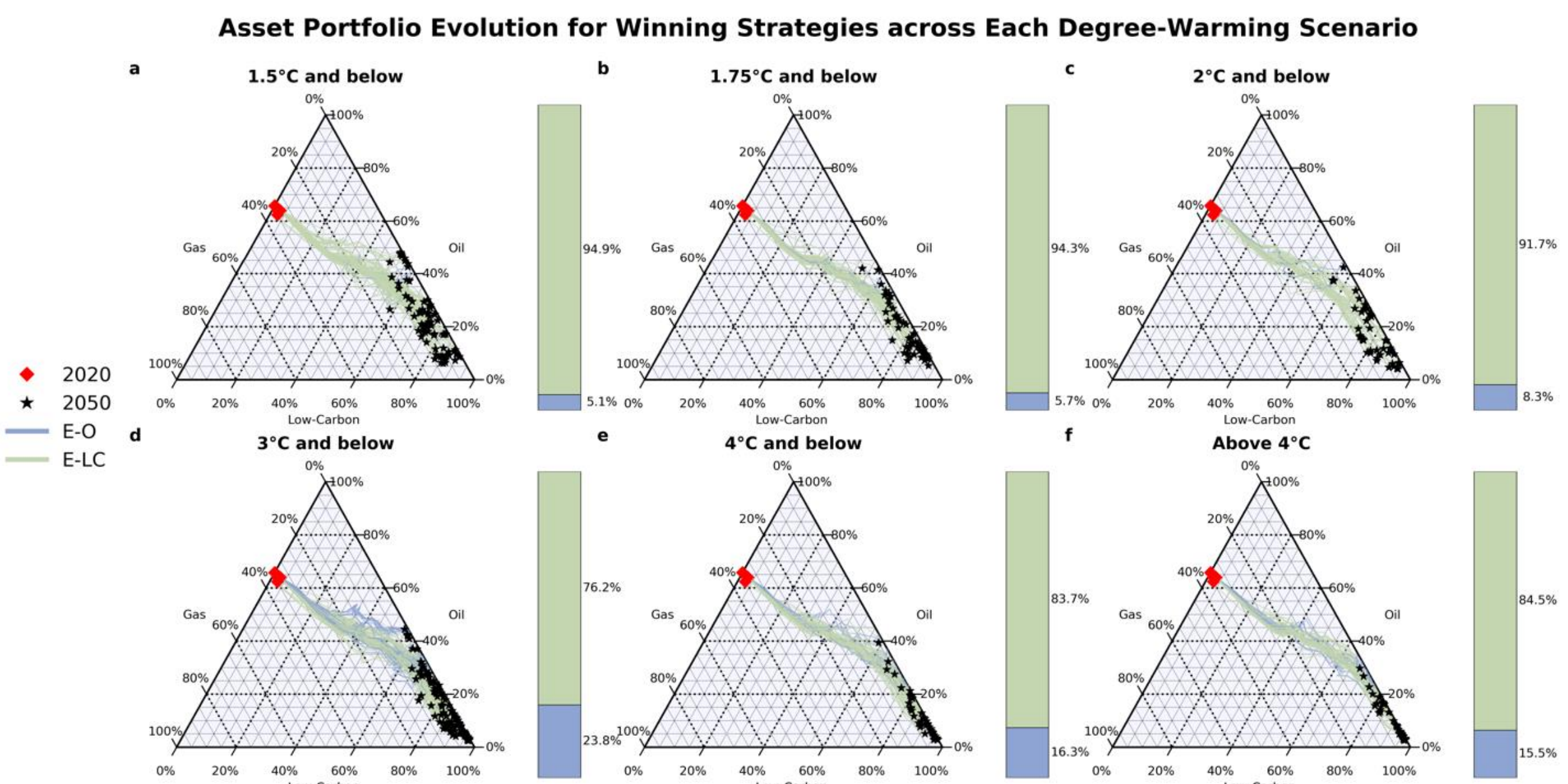}
\caption{The evolution of winning asset portfolios across each degree-warming (energy futures) scenario. \textbf{a-f},
The most winning IOC's average asset portfolio evolution across \textbf{(a)} $1.5^\circ C$ and below, \textbf{(b)} $1.75^\circ C$ and below, \textbf{(c)} $2^\circ C$
and below, \textbf{(d)} $3^\circ C$ and below, \textbf{(e)} $4^\circ C$ and below, \textbf{(f)} Above $4^\circ C$ degree-warming scenarios. Complete low-carbon
transitions by the most winning IOCs are achieved by $2028 (2024 \pm 2039)$, on average (minimum $\pm$ maximum). Bar charts to the right of each ternary graph display the distribution of winning strategy combinations for the respective degree-warming scenario.}
\label{fig:asset_portfolio_evolution}
\end{figure*}

IOCs yielding the most winning strategies - that is, leading to highest dividend payouts across each of their $21$ unique matchups - continue considerable low-carbon asset acquisitions beyond their early movement. We investigate this behavior as well as hydrocarbon tactics by observing the evolution of each \textit{IAMC/IIASA} scenarios' and the IEA's Net-Zero scenario's most winning IOCs' asset portfolios (Figure \ref{fig:asset_portfolio_evolution}).

In all cases, winning IOCs transition into low-carbon with endgame low-carbon asset
holdings accounting for $80.52\% (50.36\pm 96.08\%)$ of its asset portfolio value, on average
(minimum $\pm$ maximum). While E-O IOCs allocate most capital towards developing oil business
models, the value of their endgame reserves never exceeds the value of their low-carbon assets.
This is due to the IOC's lower cost of low-carbon asset acquisition through optimal bidding as
well as timely hydrocarbon reserve development and depletion. Focusing on the latter, we find
that IOCs exhaust their initial hydrocarbon reserves - to $16.80\% (3.63\pm 47.68\%)$ and $2.68\% (0.20\pm 15.64\%)$ of its asset portfolio, on average (minimum $\pm$ maximum), for oil and gas, respectively - rather than replenishing them. The substantial decrease from an average $65\%$ of
initial oil reserves value to $9.21\%$ supports earlier findings on the necessity of oil production to enable a robust low-carbon transition (Figure \ref{fig:emergent_strategies}h). 

Examining behavior in hydrocarbon markets, the E-LC IOCs boast the greatest performances by swiftly producing hydrocarbon assets. Oil reserve portfolio representation remains above $50\%$ until $2023 (2021-2032)$, on average (minimum $\pm$ maximum), while gas reserve representation remains above $20\%$ until $2024 (2023-2031)$, on average (minimum $\pm$ maximum). These rapid market exits are prioritized to enable larger acquisitions of low-carbon assets in the early game. Despite efforts to alleviate hydrocarbon dependency, IOCs risk stranding the carbon-intensive assets in low degree-warming scenarios (Figure \ref{fig:emergent_strategy_combinations}a-c). In these
climate-compatible scenarios, $47.5\%$ of all most winning IOCs maintain endgame oil reserves
that represent above $20\%$ of their asset portfolio. The resulting forgone revenue, due to the scenarios' lower demand and prices, limits an IOC's ability to scale its low-carbon economies.
This curb to further transition-oriented acquisitions is evidenced by the dissimilarities in endgame low-carbon asset portfolio representations amongst the most winning IOCs in lower degree-warming scenarios.

\section*{Similar strategies in diverging demand scenarios.}

\begin{figure*}
\centering
\includegraphics[width=.9\linewidth]{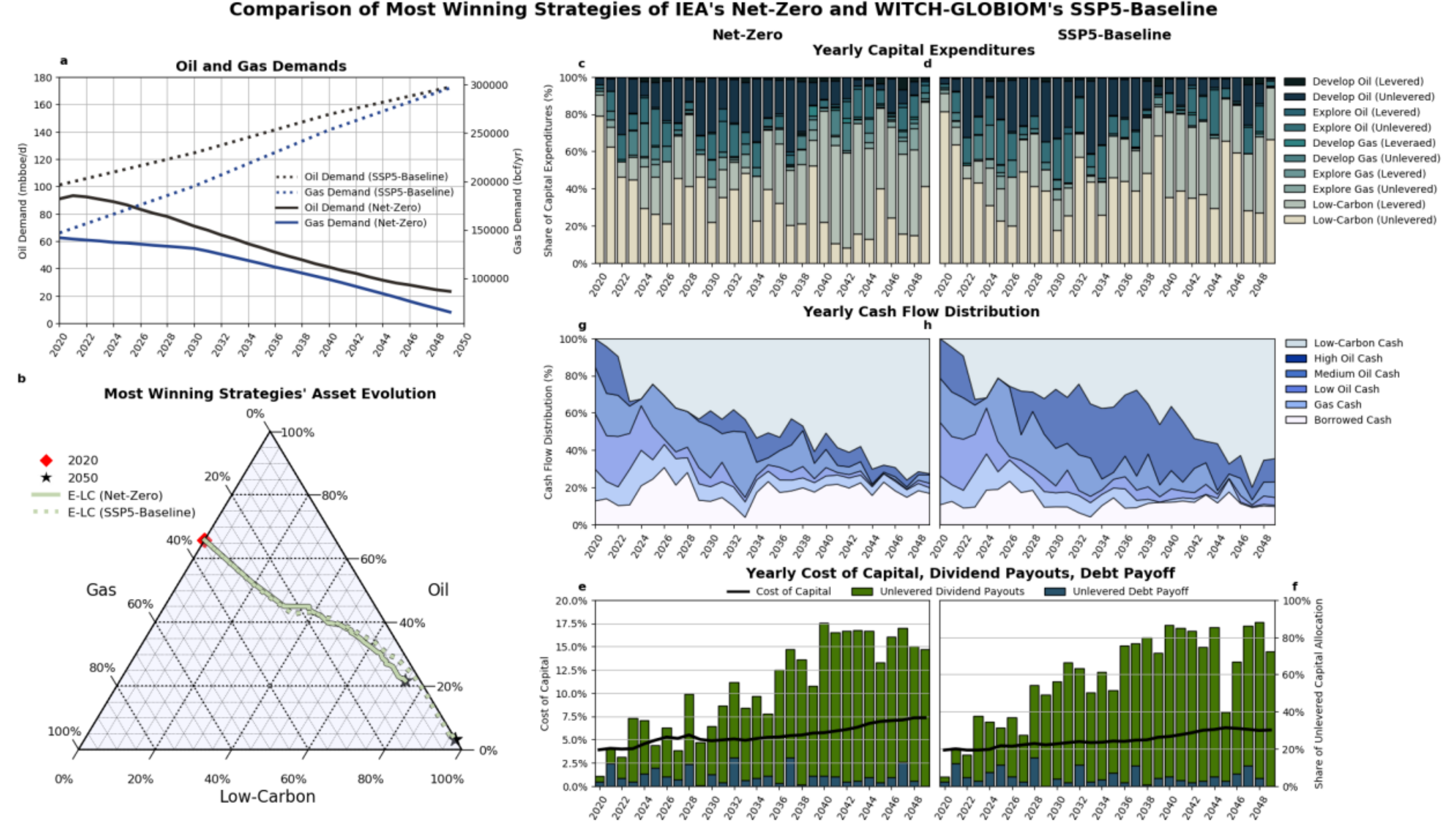}
\caption{Comparison of winning IOC's strategies in two diverging hydrocarbon demand and degree-warming scenarios. The IEA's Net-Zero and WITCH-GLOBIOM's SSP5-Baseline were chosen for comparison. \textbf{a,} Net-Zero's and SSP5-Baseline's oil and gas demand projections. \textbf{b,} asset evolution of each scenario's most winning IOC and the company's resultatn strategy combination. \textbf{c-d}, Yearly capital expenditures for the most winning IOC
in \textbf{(c)} Net-Zero, \textbf{(d)} SSP5-Baseline. \textbf{e-f}, Yearly cash flow distribution for the most winning IOC in \textbf{(e)} Net-Zero, \textbf{(f)} SSP5-Baseline. \textbf{g-h}, The most winning IOC's cost of capital as well as unlevered dividend payouts and unlevered debt payoff as a percentage of unlevered capital allocation in \textbf{(g)} Net-Zero, \textbf{(h)} SSP5-Baseline.}
\label{fig:comparison_winning_strategies}
\end{figure*}

To examine how strategies develop more closely, we compare the most winning strategies for two diverging degree-warming, hydrocarbon demand, scenarios - IEA's Net-Zero ($1.5^\circ C$ and below) and WITCH-GLOBIOM's SSP5-Baseline (Above $4^\circ C$) (Figure \ref{fig:comparison_winning_strategies}). Despite
the two scenario's asymmetric oil and gas demands, each winning IOC, on average, converges
towards a similar high-level robust strategy: scaling low-carbon business models predicated on
early ($2020$ for both) low-carbon mover behavior (Figure \ref{fig:comparison_winning_strategies}a,b). However, we find that these strategies differ with respect to four characteristics: primary financial instruments deployed to scale low-carbon economies in the late-game; year in which low-carbon cashflow becomes the main source of income; endgame hydrocarbon reserves; and dividend policies.

Observing how energy assets are acquired, or developed, we find that the most winning IOCs of the Net-Zero and SSP5-Baseline scenarios yield nearly identical capital allocation distributions (Figure \ref{fig:comparison_winning_strategies}c,d). Both IOCs boast considerable unlevered hydrocarbon development strategies yet avoid exploration activities in efforts to rapidly deplete reserves. The focus on development, therefore increasing production, allows for both IOCs to rely primarily on unlevered cashflows to finance their low-carbon acquisitions. These similar development and acquisition strategies accelerate the Net-Zero and SSP5-Baseline IOC's low-carbon transition, occurring in $2026$ and $2025$, respectively. In the last decade of play, while both IOCs decrease hydrocarbon development activities, low-carbon acquisition tactics begin to diverge as the NetZero IOC becomes increasingly reliant on raising debt to bolster its low-carbon business model. These late-game leveraged acquisitions are a result of the Net-Zero scenario's hydrocarbon demand shocks, thus diminishing respective revenues. Despite the different financial instruments deployed, IOCs continue to focus capital expenditures on low-carbon assets regardless of hydrocarbon scenario. 

The benefits of early low-carbon movement, particularly in $2020$ where low-carbon
acquisitions accounted for at least $90\%$ of capital expenditures, are seen as early as $2023$ for both IOCs (Figure \ref{fig:comparison_winning_strategies}c-h). The continued acquisition of these assets result in low-carbon returns
becoming the primary source of income for the Net-Zero and SSP5-Baseline IOCs by $2028$ and
$2039$, respectively. Though the latter exhibits a quicker low-carbon transition, as noted
previously, the SSP5-Baseline IOC continues to focus its income dependency on hydrocarbon
returns, namely heavy oil, due to the high oil demand (Figure \ref{fig:comparison_winning_strategies}h). In addition to increased cashflow, the continued production of heavier oil assets allows the SSP5-Baseline IOC to deplete its oil reserves and minimize risks of asset stranding (Figure \ref{fig:comparison_winning_strategies}b). The Net-Zero IOC, despite its rapid exit efforts, however, bears these stranded costs as it is unable to reduce its oil reserves
which represent $21\%$ of its endgame portfolio value.

To effectively build low-carbon economies, both IOCs cut dividend payouts in the earlygame and reallocate capital towards low-carbon acquisitions (Figure \ref{fig:comparison_winning_strategies}e,f). While the SSP5-Baseline IOC increases its dividends policy to allocate $50\%$ of its unlevered cashflow by $2028$, the Net-Zero IOC continues to keep dividend payouts below this level for an additional eight years.
The climate-compatible IOC follows this delayed dividends policy due to its anticipation of
lower hydrocarbon returns the following years as a result of decreased demand. Focusing
unlevered cashflows on scaling low-carbon business models enables the IOC to benefit from
stable, long-term returns. Moreover, it decreases risks of a spiraling cost of capital which allows
the IOC to raise sufficient debt to acquire low-carbon assets in the late-game without adversely affecting its bottom-line (Figure \ref{fig:comparison_winning_strategies}c,e).

\begin{figure*}
\centering
\includegraphics[width=.9\linewidth]{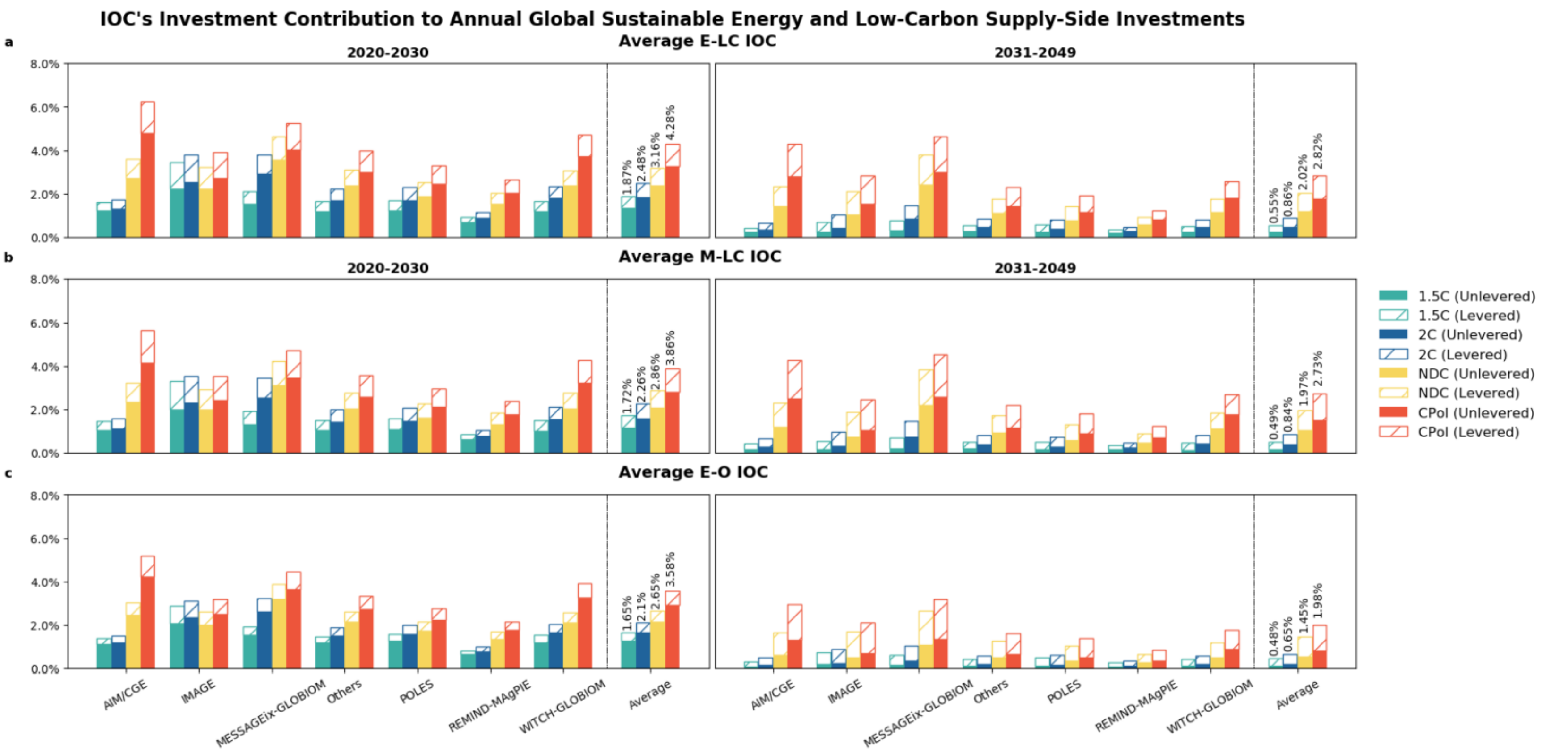}
\caption{\textbf{Emergent strategies' investment contributions to annual global low-carbon supply-side investment
across different energy models and degree-warming scenarios.} Annual global low-carbon supply-side
investment projections were drawn from a recent analysis on required low-carbon investments to achieve climatecompatible energy transitions with respect to six integrated assessment models - AIM/CGE, IMAGE, MESSAGEixGLOBIOM, POLES, REMIND-MAgPIE, and WITCH-GLOBIOM - each under four scenarios - 'Current Policies' (CPol), 'Nationally Determined Contributions' (NDCs), 'Well Bellow 2 Degress (2C)' and 'Toward 1.5 Degrees' ($1.5$C)~\citep{mccollum_energy_2018}. We allocated the required annual global low-carbon supply-side investment projections to the respective
model and scenario; NDC investment projections were allocated to $3^\circ C$ and below scenarios, CPol investment
projections to any scenario above $3^\circ C$. For remaining models, we allocated the average of all six model projections,
as well as for each of the four scenarios, grouped in the Others columns.}
\label{fig:ioc_investment_contribution}
\end{figure*}

Recently, researchers~\citep{mccollum_energy_2018} estimated the required annual low-carbon energy investment to achieve
an energy transition that would remain consistent with the goals of Nationally Determined
Contributions (NDCs) as well as of $2^\circ C$ and $1.5^\circ C$ degree-warming scenarios with respect to several IAM frameworks. We use these estimates as a proxy for 2DP's Sustainable and Low-Carbon Energy Investment Available metric, available as purchasable assets in the sustainable and low-carbon energy auction mechanism, under the respective model and degree-warming scenario to observe the potential impact our IOC's early low-carbon mover strategies have on the
global energy transition (Appendix Table A.\ref{tbl:global_scenario_metrics}).

Across all scenarios, we find the average E-LC, M-LC, and E-O IOC have the potential of playing an integral role in enabling a climate-compatible energy transition (Figure \ref{fig:ioc_investment_contribution}). This is particularly true in the $2020$s decade when global transition capital requirements begin to gain momentum and fewer potential demand shocks are present - granting IOCs stable hydrocarbon cash flows as well as credit access. On average, an E-LC IOC could provide $1.87\%$ or $2.48\%$ of required low-carbon supply-side investment to help achieve a $1.5^\circ C$ or $2^\circ C$ degree-warming scenario, respectively (Figure \ref{fig:comparison_winning_strategies}a). A M-LC IOC contributes slightly less capital at $1.72\%$ or $2.26\%$ for a $1.5^\circ C$ or $2^\circ C$ degree-warming scenario, respectively (Figure \ref{fig:comparison_winning_strategies}b). An E-O IOC allocates the least, albeit still significant, amount of capital at $1.65\%$ or $2.10\%$ for a $1.5^\circ C$ or $2^\circ C$ degree-warming scenario, respectively (Figure \ref{fig:comparison_winning_strategies}c). These figures decrease in the following decades yet remain non-negligible at $0.55\%$ or $0.85\%$ for the E-LC IOC, $0.49\%$ or $0.84\%$ for the M-LC IOC, and $0.48\%$ or $0.65\%$ for the E-O IOC in the $1.5^\circ C$ or $2^\circ C$ degree-warming scenarios. 

Instrumental to these yearly low-carbon expenditures is the access to credit markets as
highlighted previously (Figure \ref{fig:emergent_strategies}d, Figure \ref{fig:comparison_winning_strategies}c,d). Within the $2020$s decade of the $1.5^\circ C$ degreewarming scenarios, levered low-carbon acquisitions account for $29.2\%$, $33.4\%$, and $22.4\%$ of an
E-LC, M-LC, and E-O IOC's transition-oriented investments, on average, respectively. As the
temperature rises, the reliance on credit markets decreases to a minimum of $24.9\%$, $27.9\%$, and
$19.4\%$ of an E-LC, M-LC, and E-O IOC, on average, respectively. In the following two decades,
reliance on credit markets to finance low-carbon asset acquisition increases as IOCs are, then,
able to raise significant amounts of debt without adversely affecting their bottom-line, as
previously noted. With early-game dividend cuts and continued dependency on credit markets,
our findings point to the potential roles lenders and investors play in mobilizing finance to scale
the low-carbon transition and ensuring capital is responsibly allocated towards these endeavors.

\section*{Discussion}

This paper finds that IOC strategies robust to market uncertainty and adversarial, including
exploiter, strategies emerge in the form of low-carbon business models as a result of early
transition-oriented movement. Our model alleviates concerns regarding global solution
convergence guarantees by applying state-of-the-art deep multi-agent reinforcement learning
algorithms and discovering no main IOC discovers a robust business-as-usual (BAU) or delayed
transition strategy across all games evaluated (see Methods). Our results suggests that IOCs
responsibly allocating capital towards low-carbon business models could benefit from, and
accelerate, the energy transition to emerge as transition leaders. 

Observing our IOCs' emergent, robust behavior, we discover that $90.94\%$ of strategies
were predicated on scaling low-carbon business models, each resulting from movement towards
low-carbon within the first five years of play (Figure \ref{fig:emergent_strategies}). In addition to out-performing
exploitative strategies across all energy futures scenarios, transition-focused IOCs intent on early
movement towards low-carbon business models and rapid exits from hydrocarbon markets
mitigate the downside risks of oil price shocks (Figures \ref{fig:average_dividend_payout},\ref{fig:emergent_strategy_combinations}, and \ref{fig:asset_portfolio_evolution}). Comparing best-performing
strategies of two diverging hydrocarbon demand scenarios, including a Net-Zero emissions
scenario, we found each winning agent yields a similar high-level robust strategy: scaling low-carbon business models predicated on early movement towards low-carbon (Figure \ref{fig:comparison_winning_strategies}). However,
notable differences with respect to future credit market dependency, hydrocarbon production,
and investor payout policies are present (Figure \ref{fig:comparison_winning_strategies}c-f). We examine the roles these transition-oriented IOCs, as well as their lenders and investors, play in fulfilling low-carbon investment
needs to achieve climate goals. With effective engagement from their lenders and investors, IOCs responsibly reallocating of capital towards low-carbon have the potential to emerge as global low-carbon energy leaders, benefitting greater society.

As of late $2018$, oil and gas Majors spent $1.4\%$ of their capital expenditures towards
sustainable energy and low-carbon technologies in the previous decade~\citep{fletcher_beyond_2019}, on average. Although industry executives are planning on committing more capital to less carbon-intensive business models after COVID-19~\citep{dnv_turmoil_2021}, cost-cutting remains at the forefront of the IOC's agendas and they are more willing to maintain, or increase, capital expenditures in the year ahead as compared to after the $2014$ price crash. We find that it is of utmost importance to allocate such capital to scale low-carbon business models, as opposed to oil and gas exploration, activities to best maximize value and mitigate transition risks, namely asset stranding and low hydrocarbon prices reducing dividend allocations. Focusing on building low-carbon economies, with the sufficient leverage, situates IOCs in more robust financial positions than those who delay low-carbon movement as well as those who continue BAU practices in efforts to exploit possible future gaps in hydrocarbon markets. Our results show that the emergent transition-focused strategies outperform hydrocarbon-centric strategies as well as effectively balance returns to investors and ease debt pressures throughout a range of oil prices.

While climate pundits point to IOCs' exposure to transition risks and incompatibility with climate-aligned scenarios~\citep{carbon_tracker_fault_2020}, industry attempts to justify their hydrocarbon-focused strategies with increased demand projections. Our results show that, regardless of demand scenario, immediately reallocating capital from hydrocarbon reserve replenishment (exploration) and into low-carbon business models boasts a more favorable strategy in satisfying investors and mitigating debt engulfment. We do not expect cuts to hydrocarbon production, however, capitalizing on returns from current reserves would raise low-carbon spending as well as reduce the amount of assets susceptible to stranding. Rather, we find hydrocarbon production becomes gradually less relevant to a company's cashflow. This pace is typically hastened in demand shock scenarios. The financial risks of pursuing such transition-oriented strategies, however, could create high financial risks, namely escalating cost of capital, as efforts to achieve climatecompatible economies (e.g. lower demand for hydrocarbons) accelerate. Our results demonstrate that immediately reallocating unlevered cashflows from hydrocarbon production mitigates such risks, even as raising debt to finance future low-carbon acquisitions becomes necessary. Thus, it is imperative for IOCs to adopt strategies centered on early movement towards a low-carbon transition before the financial implications of transition risks are felt and magnified.

Lastly, our findings point towards the mutually beneficial outcomes of low-carbon
movement shared between the transition-oriented IOC and the global energy transition. Despite
reaching a new high in $2020$ with over $\$500$B~\citep{bloombergnef_energy_2021}, current global low-carbon energy investment
trends still fall short in attaining a climate-compatible energy transition~\citep{mccollum_energy_2018}. IOCs adopting low-carbon strategies could help fill these investment gaps, especially in the current decade where transition-focused finance needs to take center stage~\citep{caldecott_defining_2022}. We note this impact, however, ultimately relies on the abilities of lenders to mobilize transition-oriented finance and investors engaging with IOCs to responsibly reallocate capital towards climate-compatible business models.

\section*{Conclusion}

Our model reveals that IOCs have the potential to overcome uncertainty by emerging as lowcarbon energy leaders. To achieve this and mitigate downside risks, lenders and investors should effectively engage with IOCs to ensure responsible reallocation of capital that would enable timely shifts towards robust, low-carbon business models. Doing so will require IOCs to increase disclosure efforts regarding their hydrocarbon and low-carbon activities as well as understand their potential in enabling a global energy transition. While it is important to note that our model does not directly restrict hydrocarbon production through emissions targets, our findings reinforce the low-carbon consensus that IOCs must transition to maximize value and mitigate downside risks regardless of degree-warming scenario. To this end, we argue IOCs can, and should, emerge as low-carbon leaders to benefit its stakeholders as well as greater society.

This work provides a data-driven analysis to support literature's low-carbon consensus
that IOCs can and should transition. Our model complements energy futures research and applied
game theory in many ways. From a broad view, the integration of deep multi-agent
reinforcement learning addresses non-linearities inherent to IAM scenarios as well as opens
opportunities for new ways of assessing uncertainty in energy scenario analysis. Specifically, our model allows for decision-makers to stress test core governance questions as well as explore the
emergent behavior of intelligent agents exposed to a range of market designs. Oil and gas
literature could bolster their qualitative findings by adding a game-theoretic evaluation to quantify risks and rewards of specific strategies with respect to a chosen energy future(s). The multi-agent league system could incorporate other competing entities to further assess robustness of an IOC's optimized strategy.

\section*{Methods}

\subsection*{2 Degree Pathways Wargame}

The 2 Degree Pathways (2DP) wargame is a decision support tool developed by the Oxford Sustainable Finance Programme and E3G, a think tank, to ``help inform company, investor, government and civil society thinking around the pathways the oil and gas majors can take to become $1.5^\circ C$/$2^\circ C$-compatible'' by simulating oil companies in competition within varying transition scenarios~\citep{tomlinson_crude_2018}. The tool was designed based on a range of wargaming literature and ex-oil and gas executive input. Two versions of 2DP were developed, one with IOCs and the other with the addition of NOCs. This work builds on the former. Wargame participants role play as fictious IOCs, bearing resemblance to real IOCs, with the goal of maximizing shareholder value by responsibly allocating capital on a year-by-year basis. Players participate in eleven markets in 2DP: two exogeneous oil and gas markets, with demand driven by the scenario at-play, and nine endogenous markets comprised only of game players, one for each for balance sheet asset. Game actions are continuous giving the game a high-dimensional state and action space. Details on this work's 2DP variation, such as player setup and game stages, are further detailed below.

\subsubsection*{Player Setup} Players maintain eleven on-hand asset classes and sixteen pipeline assets, seven decision-making metrics and the ability to choose from $64$ actions (Appendix Tables A.\ref{tbl:ioc_balance_sheet_distribution}, A.\ref{tbl:ioc_decision_making_metrics}, A.\ref{tbl:ioc_action_space1}, and \ref{tbl:ioc_action_space2}, respectively). All players yield the same oil and gas capital costs (Appendix Table A.\ref{tbl:hydrocarbon_capital_costs}). While equal in total cash value, players begin with differing asset distributions to represent how an initial market dominance, or diversity, may affect strength of overall strategies (Appendix Table A.\ref{tbl:ioc_balance_sheet_distribution}). These values were calculated based on the average respective asset holdings seen in six oil Majors' annual reports ~\citep{sp_sp_2020} to best represent market conditions. A player's level of assets accumulated and actions chosen throughout the game make up the respective player's decision-making metrics for a given year (Appendix Table A.\ref{tbl:ioc_decision_making_metrics}).

\subsubsection*{Scenario Setup} Global scenario metrics dictate the game environment's dynamics, asset payoffs, and a player's strategy. Upon game initiaion, beginning in year $2020$, these metrics are taken from the selected energy futures scenario, where each year of the scenario's data represents the metrics for the respective in-game year. In this work, we acquire data to represent $408$ energy futures scenarios from the Integrated Assessment Model Community (IAMC) and International
Institute for Applied Systems Analysis (IIASA) $1.5^\circ C$ explorer database, IEA's Net-Zero
report~\citep{iea_net_2021}, \citet{mccollum_energy_2018}, OPEC historical trends~\citep{opec_opec_2020}, as well as BP~\citep{parnell_how_2020} and Equinor~\citep{equinor_presenting_2021} statements
(Appendix Table A.\ref{tbl:ioc_decision_making_metrics}). Given multiple scenarios to test across, we incorporated a random energy future scenario generator to prevent agents from overfitting to a single scenario's demand curve. Moreover, they are unable to play the same transition scenario twice without having played through each of the other scenarios at least once. At the end of each game, year $2050$, the game resets and the generators selects a new scenario for the following epoch. 

\subsubsection*{Game Stages} 2DP's model breaks one year of play into four transaction stages: production, borrowing, trading, and allocation. Each stage is explored further below and their respective player actions found in Appendix Table A.\ref{tbl:ioc_action_space1}, and \ref{tbl:ioc_action_space2}.

\paragraph{Production} The production stage calculates the player's net income for a given year. Net income include cash gained from oil production, gas production, low-carbon assets, and cashj used to pay off debt interest, if any. A key feature to this stage is oil price formation predicated on the current scenario's oil and gas demand, OPEC \& other's production share global metric, and the sum of produced oil and gas assets players choose to produce in a given year. Therefore, players must appropriately produce oil and gas assets that provide sufficient returns each year as overproducing (underproducing) may leadto glut (missed returns). Returns from low-carbon assets are predicated on the current year's sustainable energy and low-carbon asset return to investment.

\paragraph{Borrowing} The borrowing stage allows a player to borrow acquire an amount of credit
dependent on the health of their balance sheet. Players are only able to borrow additional cash if
their current debt-to-equity ratio is below $200\%$. Player's chosen borrowed amount is added to their cash assets and debt liabilities.

\paragraph{Trading} Trading is split into two sub-mechanisms: one sustainable energy and low-carbon investment auction and one player-to-player trading platform. In the former, players simulate inorganic low-carbon growth by placing bids to purchase the respective assets in sealed-bid auction form. The low-carbon auction allows players to bid with cash and credit separately, designed to further explore transition finance behavior with respect to the two endogenous markets. While submitted independently, all cash and credit bids compete for the same amount of sustainable energy and low-carbon assets available of a given year. These auctioned assets are dictated by the scenario at hand (see Appendix Table A.\ref{tbl:global_scenario_metrics}). Auction sales follow the order of the highest bid placed. A player with the highest bid maintains purchasing priority. Players with lower bids are at risk of being unable to purchase low-carbon assets due to the finite amount of assets held within the bank. 

Post-auction, players have the choice to participate in trading amongst each other. All
hydrocarbon and low-carbon on-hand assets are available to trade. Similar to the above, sale of assets follow the order of the highest bid placed for the respective asset. Multiple sales from the same player to others may occur if the volume of an asset up for sale is greater than the other players' bidding volume for said asset. No trades occur if there is no buyer or no seller.

\paragraph{Allocation} The allocation stage grants players the ability to explore and/or develop oil and/or gas assets, pay off debt, pay dividends, and/or save cash into the next year. Similar to the lowcarbon auction, allocation actions are split between credit-only and cash-only actions. The motivation for this is to restrict agents from borrowing significant amounts of cash only to pay as dividends, an unrealistic business model.

\subsection*{Game Theory}

This work's focus on revealing robust strategies to energy scenario uncertainty and adversarial entities is a game theoretic problem. Game theory is the study of strategic interactions between a set of agents, or players, in game form~\citep{von_neumann_theory_1944}. Games are largely categorized by how its agents' total losses and total gains are summed. We frame 2DP as a non-zero (general-) sum game - that is, a game in which strategic payoffs may sum to values greater than, or less than, zero - by centering each agent's utility payoff on its allocated total dividends. Agents are said to be in Nash Equilibrium when strategy profiles converge to a point in which they cannot be improved upon by unilateral deviation~\citep{nash_equilibrium_1950}. Of particular importance of this work is the concept of approximate mixed-strategy Nash equilibria whereby at least one player is playing a non-deterministic strategy and the conditions of a Nash equilibria are \textit{approximately} satisfied. 

Computing Nash equilibria of n-player, general-sum games using conventional game theoretic methods is impractical~\citep{shoham_multiagent_2008}. While some problem formulation possibilities exits (e.g. approximating Nash using a sequence of linear complementarity problems approach or reverse the problem to optimize for a minimum) as well as applicable algorithms (e.g. n-player extensions of Scarf, Lewke-Howson, and the support-enumeration method), no method exists that could effectively solve for an infinitely-repeated game, such as 2DP. Moreover, modeling the 2DP wargame's non-linear payoffs as well as high dimensional state and action spaces proves analytically and computationally intractable.

\subsection*{Deep Reinforcement Learning}

We use reinforcement learning, the computational approach to learning from environment interaction~\citep{sutton_reinforcement_2018}, to have our IOC agents learn robust strategies under energy futures uncertainty. Reinforcement learning algorithms seek to solve problems defined as a Markov Decision Processes (MDPs) comprising of four key components: the environment and action space, the observation space, describing the game elements accessible to an agent, and reward function. Stochastic, or Markov, games extend MDPs to involve multiple agents whose actions impact their resulting rewards and next state. To best mimic real-life market competition, we frame the 2DP wargame continuous control problem as a partially observable stochastic game (POSG) which restricts an agent's observation space to a limited set of information. POSGs rely on the same four general components used in MDPs. This work utilizes the aforementioned variant of 2DP and its player's available actions as the game environment and action space. Agent state observations, following the POSG framework, are incomplete and imperfect (i.e. restricted to limited game and opponent information) to best mimic the partial observability of real-world market competition (Appendix Table A.\ref{tbl:ioc_game_state_observations} The agent's reward function, $r$,focuses on maximizing shareholder value via dividend payouts and debt engulfment mitigation (Appendix Table A.\ref{tbl:ioc_reward_function}).

The incorporation of the aforementioned reinforcement learning components make up each agent's discounted value function $V_\pi(s)$ and policy function $\pi_\theta(a|s)$. The former is used to predict rewards in a given state $s$,typically discounted by some factor $\gamma$ in infinite horizon games, such as 2DP, while the latter represents the strategy an agent takes by playing action a for a given state observation $z$. In the context of this work, we follow the actor-critic setting~\citep{konda_actor-critic_1999}. Here, the agent's actor seeks to converge at an optimal policy function $\pi_*(a|s)$ by finding parameters $\theta$ that maximize the performance measure $J(\theta)$, some loss objective function, of a policy through gradient descent. The agent's critic, where learning takes place, computes a value function $V_\pi(s)$ to best predict $r$ by critiquing the actor's policy function updates. These updates are applied as the agent generates new trajectories (i.e. training data) at every game time-step (i.e. year of 2DP play).

A challenging task for reinforcement learning is solving an environment of highdimensional continuous state and action spaces, such as the 2DP wargame. The introduction of artificial neural networks as function approximators enables deep reinforcement learning algorithms to effectively address this challenge. In this work, we apply a combination of neural networks and actor-critic based reinforcement learning methods, detailed below, to improve agent performance.

\subsubsection*{Algorithm} We train agents with the advantage actor-critic~\citep{wu_openai_2017} (A2C) and proximal policy
optimization (PPO) algorithms~\citep{schulman_proximal_2017}, equipped with deep neural networks, due to their state-of-the-art performance on high-dimensional continuous control tasks~\citep{openai_dota_2019,schulman_proximal_2017}.

A2C follows the actor-critic method as described above but replaces the value function with an advantage function, measuring the improvement of taking an action $a_t$ over the average action $\overline{a}_t$ in that state, to increase performance. Moreover, it enables the agent to engage in continuous state and action spaces by modeling $\pi_\theta(a|s)$ as a Gaussian distribution. In this work, we equip the actor and the critic with separate deep neural networks to serve as policy and value function approximators, respectively, allowing for the effective mapping of state-action pairs to expected rewards. A2C alone, however, is insufficient to solve a high-dimensional continuous control task due to its high sensitivity to hyperparameter tuning and susceptibility to training instability due to large policy updates.

Built on A2C, PPO responds to the issues evident in the actor-critic method, as well as in past policy-based learning methods, by modifying the policy update performed in the actor network with a more conservative policy update region via a clipped surrogate objective function $J^{\text{CLIP}}(\theta)$. With PPO, policy updates are performed by collecting a batch of trajectories (i.e. experiences from game play) and are optimized with mini-batch stochastic gradient descent to increase stability and convergence. Additionally, PPO addresses concerns regarding large number of samples and high variance present in high-dimensional continuous control problems by upgrading A2C's advantage function with a generalized advantage estimator.

\subsubsection*{Agent Training} We use an independent learning (InRL) training method to decompose an $n$-agent multi-agent reinforcement learning (MARL) problem to a single learning agent problem whereby other (frozen) agents are treated as part of the localized environment. The application of InRL allows for the implementation of the 2DP league mechanism described below. Concerns of this InRL approach, however, arise with respect to its theoretical limitations that would result in learning instabilities, overfitting, and loss of convergence guarantees~\citep{lanctot_unified_2017,laurent_world_2011}. Despite these potential issues, the use of PPO, specifically the multi-agent variant Independent PPO, in InRL problems has been shown to match, and even outperform, state-of-the-art MARL algorithms in multi-agent settings~\citep{schroeder_de_witt_is_2020}.

\subsection*{Multi-Agent Learning}

To address game-theoretic challenges and encourage more robust agent strategies during training, we introduce league training similar to the AlphaStar League~\citep{vinyals_grandmaster_2019}. Central to our 2DP-League training setup is the incorporation of self-play and prioritized fictitious self-play (PFSP) variants. We iteratively populate the league with players that represent saved parameters from previously trained main and exploiter agents. 

\subsubsection*{Self-Play} Self-play~\citep{hernandez_generalized_2019} is a training mechanism central to multi-agent learning. In self-play, a learning agent is trained by competing against itself. This mirrored battle provides sufficient amount of challenge such that an agent achieves superhuman-level performance and complex emergent behavior~\citep{tesauro_temporal_1995,silver_mastering_2016,vinyals_grandmaster_2019,openai_dota_2019,bansal_emergent_2017}. Our self-play algorithm updates all agents, learning and frozen, in the environment with the learning agent's latest policy update every epoch so long as it is a winning policy (i.e. win-rate greater than or equal to $50\%$). While the same policy, an agent's policy output may differ from its opponents' policy outputs as we train agents playing mixed strategies to encourage convergence towards a mixed strategy Nash equilibrium. The motivation for this is the thinking that, for example, company A should not, and cannot, expect company $B$ to play an expected, deterministic strategy. Training against deterministic policies would result in company $A$ overfitting its strategies towards a narrowed belief in its opponents' strategies.

\subsubsection*{Prioritized Fictitious Self-Play} Overfitting is of concern with respect to multi-agent learning algorithms such as self-play~\citep{bansal_emergent_2017}. Fictious self-play (FSP)75 helps prevent this issue and the
occurrence of strategy cycles by uniformly sampling from previous saved policies. Sampling
from all previously saved policies, however, is compute-intensive and inefficient as many games
would be played against extremely poor policies. To combat this inefficiency, AlphaStar
proposed PFSP whereby a matchmaking system replaces the uniform sampling to grant a
learning signal. In this work, we employ tow variations of AlphaStar's PFSP, PFSP-past and
PSFP-opponent. We sample a frozen agent (PFSP-past), or group (PFSP-opponent), from frozen
policy pool $C$ for the learning agent to play with the probability
\[
    \frac{e^{f_{\text{weight}}}}{\sum_{C\in\mathbf{C}}e^{f_\text{weight}}}
\]
where $f_{\text{weight}}$ is some weighting function described by the PFSP mechanism employed. 

In PFSP-past, we sample from past-selves of the learning agents initialized using the weighting function $f_{\text{hard}}(x)=\left(1-x\right)^2$, where $x$ is the win-rate of the learning agent, to encourage play against the most difficult past-selves. We iterate through this sampling $N-1$ times, where $N$ represent the number of total players in a match. In the event the learning agent fails to learn (i.e. unable to beat the frozen agents for more than several consecutive epochs), we reset the matchmaking mechanism and use $f_{\text{var}}(x)=x\left(1-x\right)$ as the new weighting function to encourage play against frozen agents around the learning agent's skill. Once learning has been sufficiently facilitated, we reset the matchmaking mechansim again to $f_{\text{hard}}$.

When playing against opponent agents, we initiate the league with unique combinations of opposing player policies from the latest training iteration. We do not populate the league with all past opposing player polices as involving all such policies would the number of unique combinations exponentially, requiring unattainable amount of compute. These combinations' weights are not initialized. Instead, the learning agent plays each combination once before allocating each $f_{\text{hard}}$ and prioritizing the most difficult combinations. Like in PFSP-past, if the learning agent is struggling, we reset the matchmaking mechanism such that all opponents have an equal chance of being played, and back to $f_{\text{hard}}$ once a learning signal is created.

\subsubsection*{League Participants} 

We employ two distinct agents that differ in the evolving distribution of
opponents they play against, when training parameters are reset, and the constraint on actions
available.

Main agents train as IOCs with no constraints on their available actions yet differ in their
initiated asset levels. In the first training iteration, main agents are trained with probability of
$80\%$ self-play and $20\%$ PFSP-past. The PFSP-opponent mechanism is introduced in subsequent
iterations beginning at $35\%$, with $50\%$ self-play and $15\%$ PFSP-past, and scaling linearly by
iteration to a maximum of $100\%$. Main agents' parameters never reset. The trained policy at the end of each iteration is saved to later be used as the based parameter for the respectivemain agent's next iteration, facilitating continuous learning. end of each iteration is saved to later be used as the base parameter for the respective main agent's next iteration, facilitating continuous learning.

Exploiter agents are initiated with oil-dominant initial assets yet are constrained to a
select group of actions based on their pre-determined strategy type. Exploiter agents are always
trained with $100\%$ PFSP-opponent against the main agents. The idea is to exploit weaknesses in the main agents' strategies, thus making them robust to both energy scenario and opponent uncertainty. The exploiter agents' trained parameters are added to the league after each training iteration. Their parameters are reset to encourage diversity in exploitation.

\subsection*{Evaluation}

Agents were evaluated across all scenarios for all different combinations of opposing agents. We evaluated eight agents - six main and two exploiters - in a six-player game across $408$ energy scenarios, totaling $11,424$ unique games (each agent plays $8,568$ games). Selected main agent policies were drawn from their respective agent's latest saved policy. Exploiter agent policies were sampled from the league. Evaluating more agents in league, as noted previously, in this way would require significant levels of compute.

\bibliography{main.bib}

\pagebreak

\onecolumn

\appendix
\section*{Appendix A.1 - Energy scenarios found in the IAMC/IIASA ensemble~\citep{huppmann_iamc_2018}}


\begin{table*}[!htb]
\centering
\includegraphics[width=.9\linewidth]{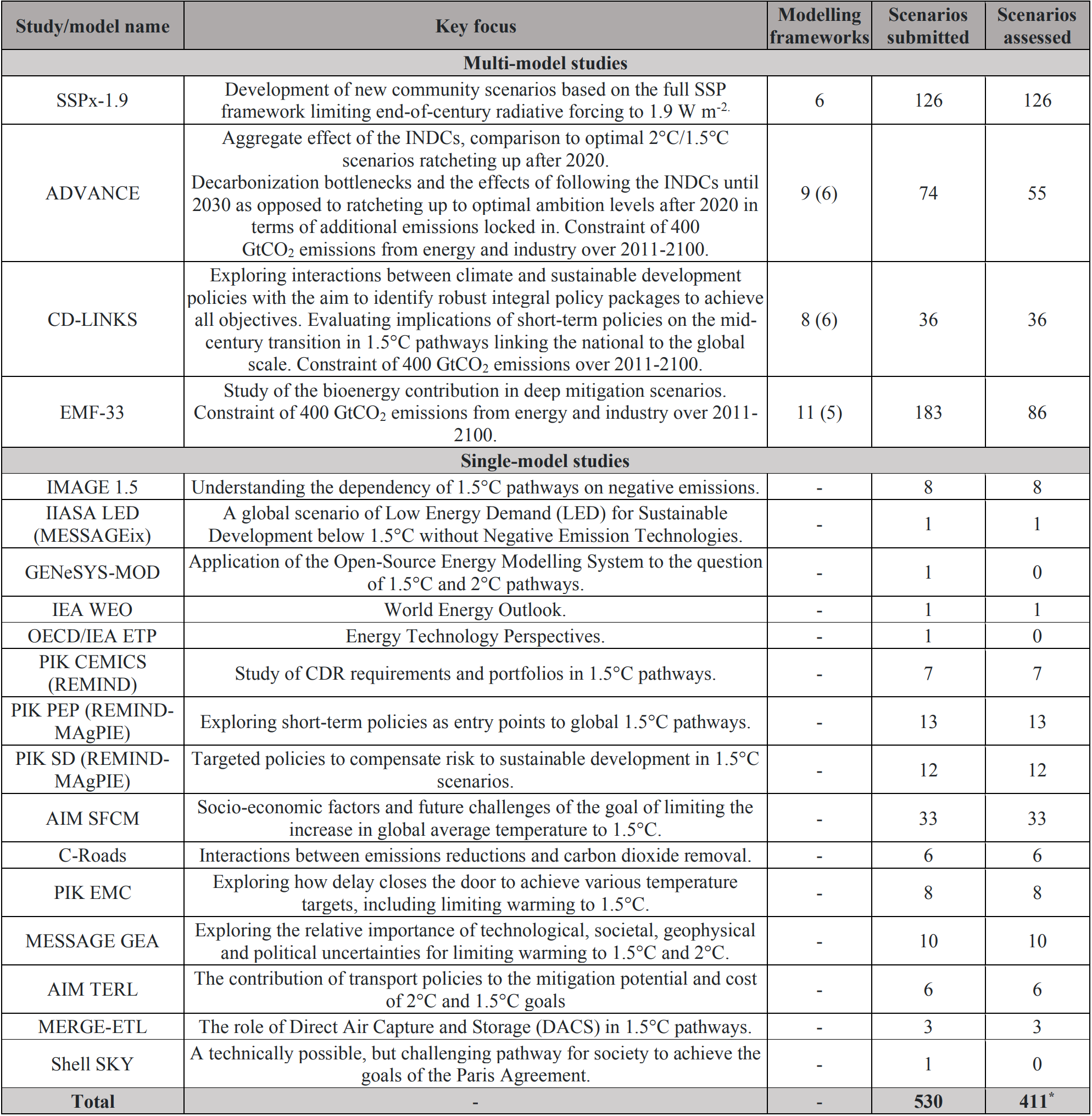}
\caption{$^*$* Though $411$ scenarios were assessed, only $409$ scenarios provided oil and gas demand projects. Two of these scenarios boasted unrealistic, outlier oil and gas demand projects (starting at approximately $30\%$ of current demand and falling)}
\label{tbl:energy_scenarios}
\end{table*}

\raggedbottom
\pagebreak

\section*{Appendix A.2 - Global scenario metrics}

\begin{table*}[!htb]
\centering
\includegraphics[width=0.7452\linewidth]{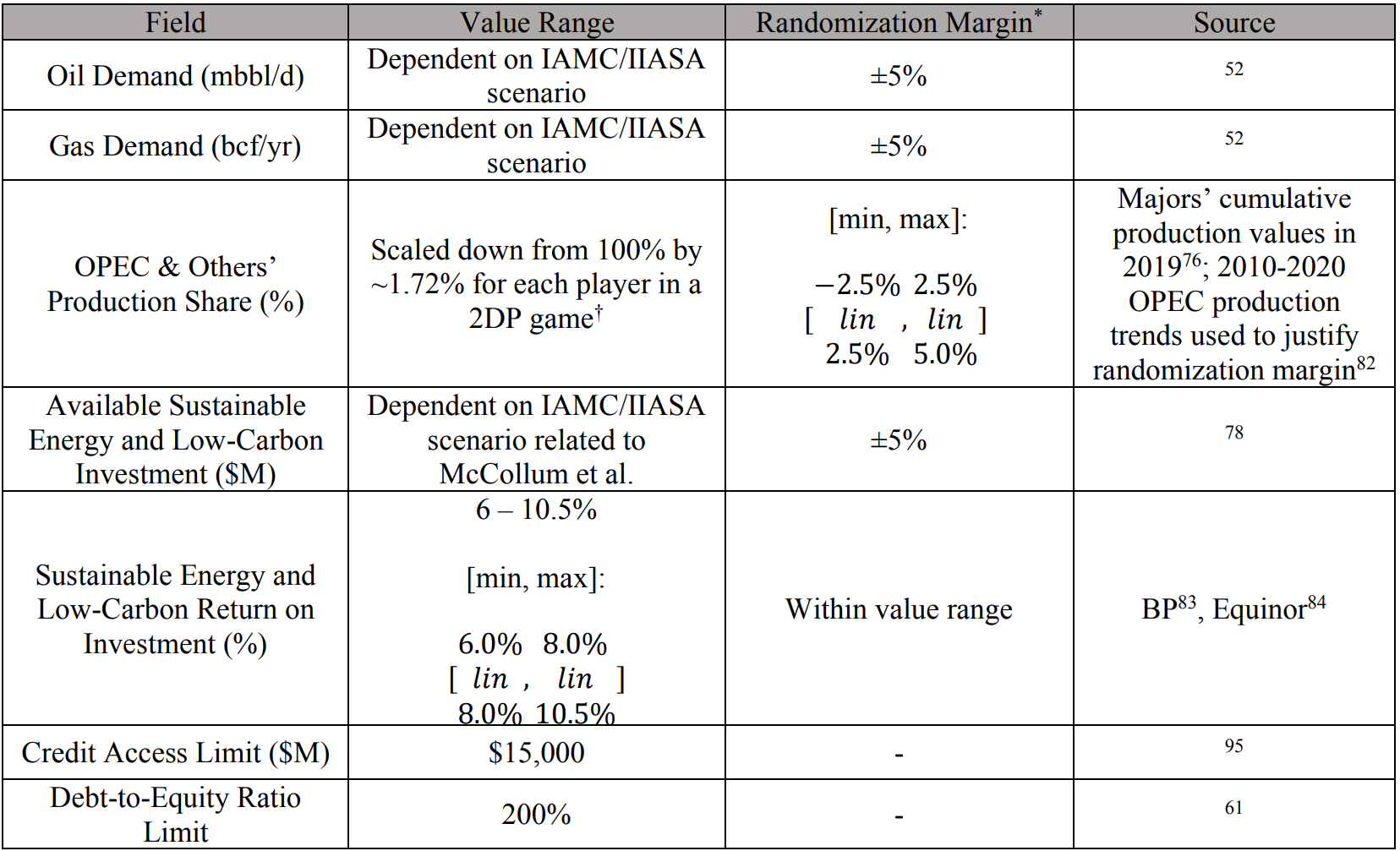}
\caption{$^*$Randomization margin represents the degree to which a value may deviate from its original value. This was introduced for several reasons: to prevent agents from overfitting to similar demand curves, prevent agents from learning to optimize with respect to the endgame (i.e. play as though it were a finite repeated game), encourage agents to discover further robust strategies with respect to any, reasonable scenario metrics.
$\quad^\dagger$ The OPEC \& Others' Production Share played a considerable role in dictating our agents' strategies.
It is commonly noted~\citep{damodaran_cost_2021,stevens_oil_2012} that future oil prices are uncertain as trajectory is hihgly dependent on a given day's geopoliotical landscape - the OPEC influence and dilemma. Rather than suggest OPEC \& Others' production levels will follow a general trendline,  we attributed the average production seen across the six oil Majors with respect to global production as the baseline production ratio for each agent. 
This way agents have equal hydrocarbon production ratios throughout the game. We introduce noise to this OPEC \& Others' Production Share metric that essentially increase, or decreases, an agent's available hydrocarbon production as well as the global oil and gas prices. 
The idea is to make our agents' strategies more robust with respect to an oil cartel and geopolitical uncertainty as seen in the real world.
}
\label{tbl:global_scenario_metrics}
\end{table*}

\section*{Appendix A.3 - IOC game state observations}

\begin{table*}[!htb]
\centering
\includegraphics[width=0.76365\linewidth]{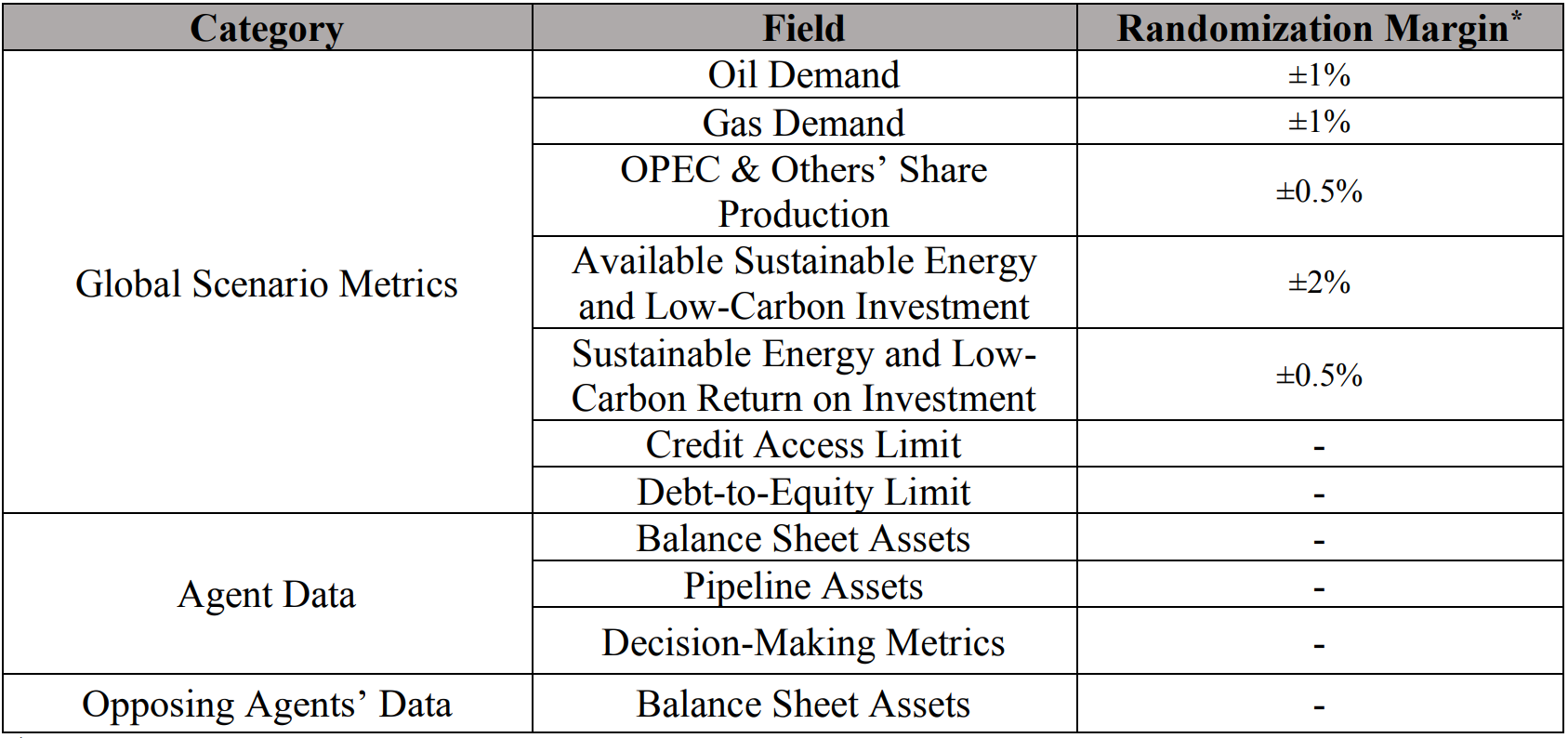}
\caption{$^*$ Further randomization was introduced into agent observations directly with the purpose of preventing agents from solving for the endgame, and instead play a game with an infinite horizon. The randomization here holds the same for each agent and, unlike the randomization introduced in Appendix Table A.\ref{tbl:global_scenario_metrics}, does not affect the actual metric (e.g. oil demand) it is imposed upon.}
\label{tbl:ioc_game_state_observations}
\end{table*}

\pagebreak

\section*{Apendix A.4 - IOC action space categorized by the game stage in which they are applicable}

\begin{table*}[!htb]
\centering
\includegraphics[width=0.8955\linewidth]{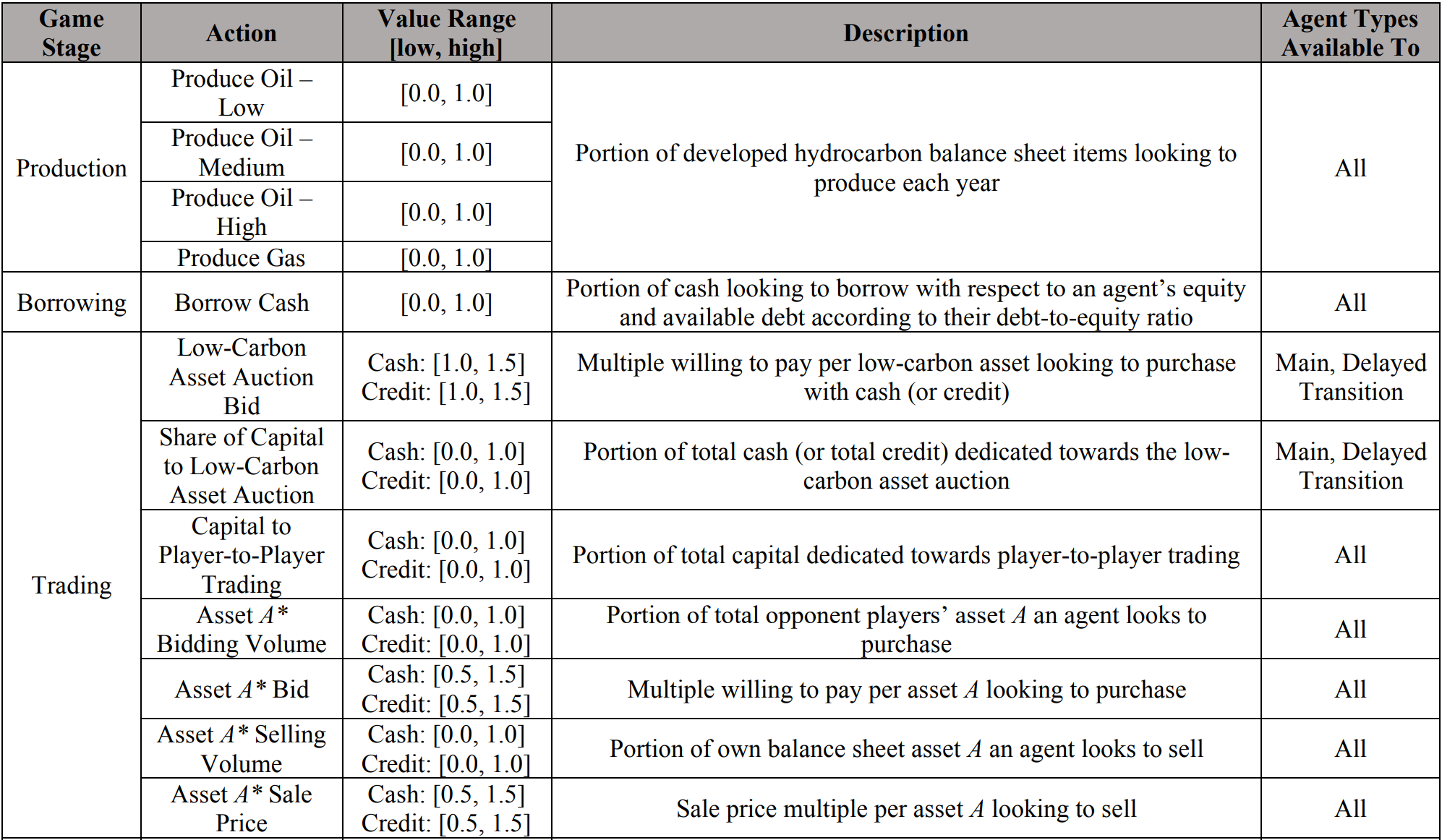}
\caption{IOC action space categorized by the game stage in which they are applicable. IOC action space categorized by the game stage in which they are applicable. $^*$ Asset A represents any hydrocarbon or low-carbon balance sheet item. In total, there are nine balance sheet items available to trade.}
\label{tbl:ioc_action_space1}
\end{table*}

\begin{table*}[!htb]
\centering
\includegraphics[width=.9\linewidth]{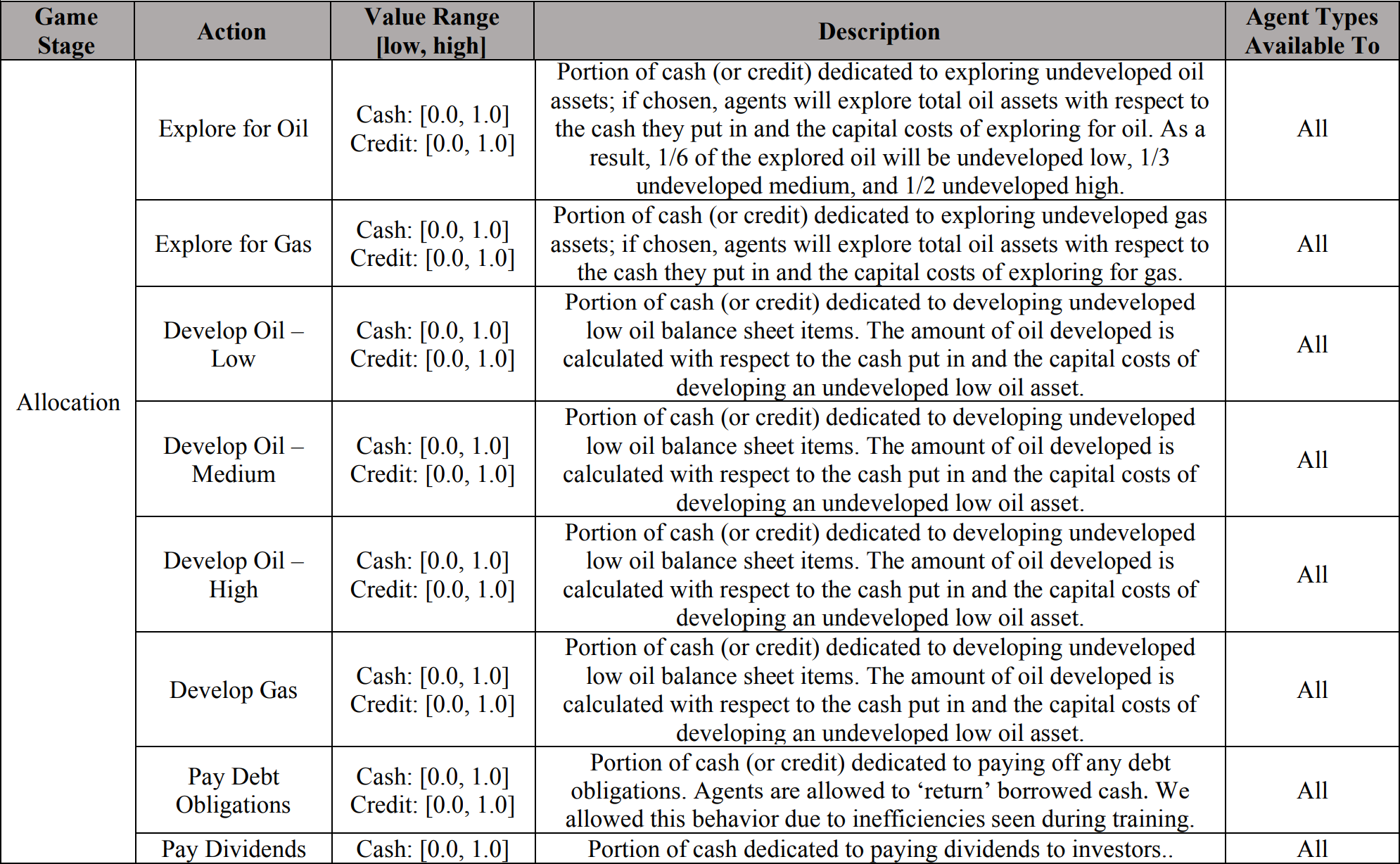}
\caption{IOC action space categorized by the game stage in which they are applicable (cont.). }
\label{tbl:ioc_action_space2}
\end{table*}

\section*{Appendix A.5 - Reward function}

\begin{table*}[!htb]
\centering
\includegraphics[width=0.75195\linewidth]{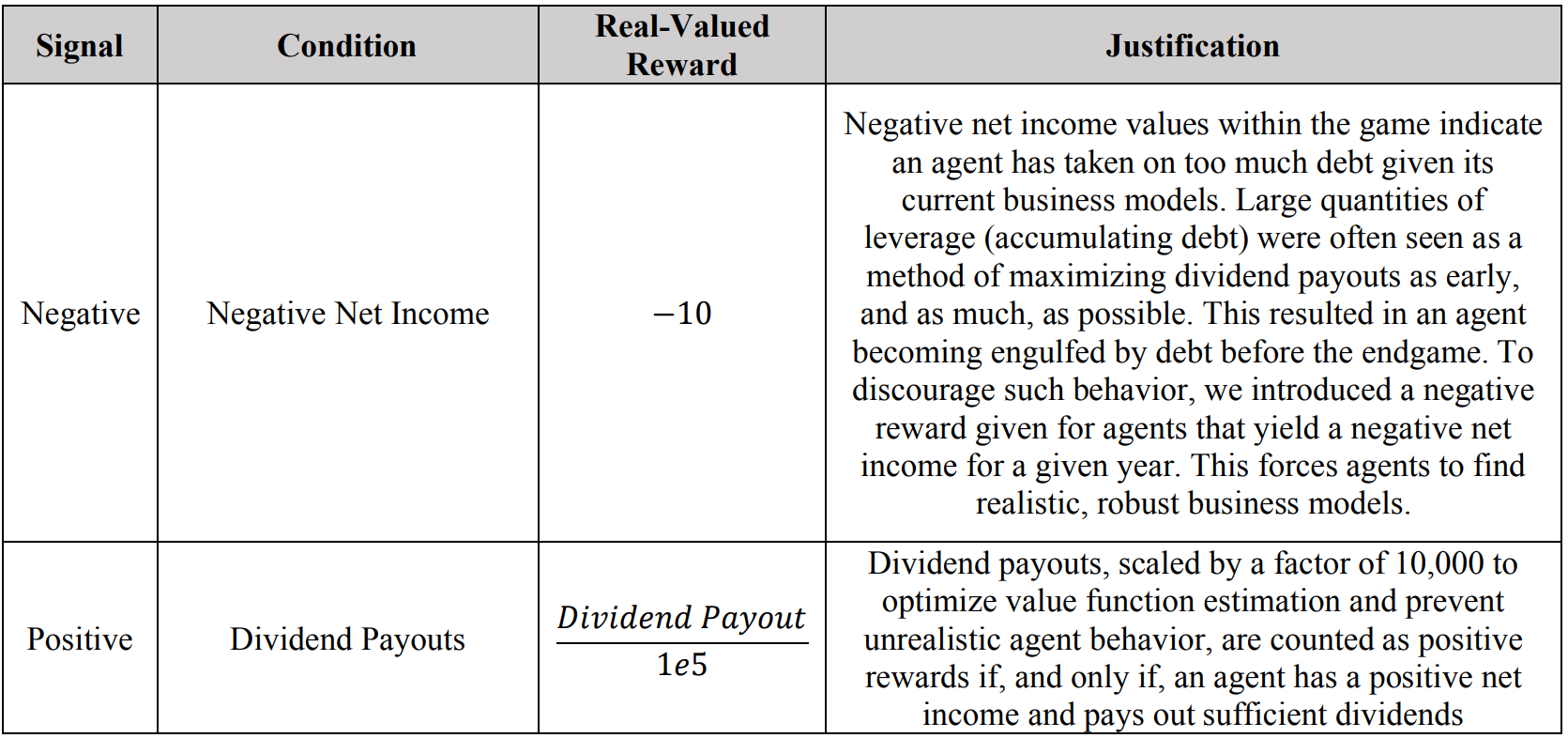}
\caption{IOC action space categorized by the game stage in which they are applicable (cont.). }
\label{tbl:ioc_reward_function}
\end{table*}

\section*{Appendix A.6 - Hydrocarbon capital costs}

\begin{table*}[!htb]
\centering
\includegraphics[width=0.46305\linewidth]{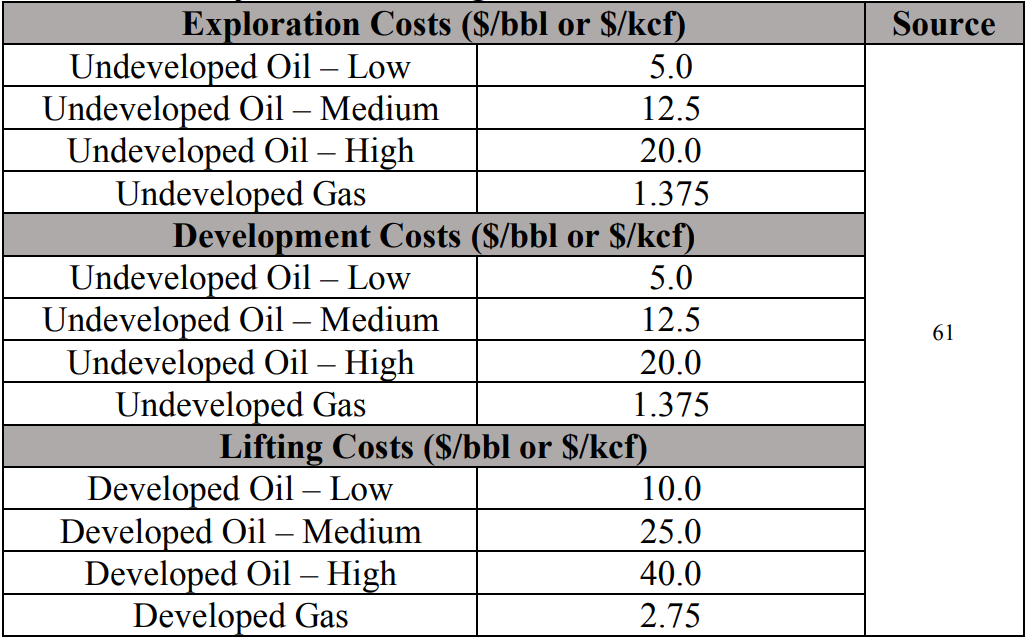}
\caption{ }
\label{tbl:hydrocarbon_capital_costs}
\end{table*}

\pagebreak

\section*{Appendix A.7 - Balance sheet distributions used as initial conditions for respective IOCs}

\begin{table*}[!htb]
\centering
\includegraphics[width=.9\linewidth]{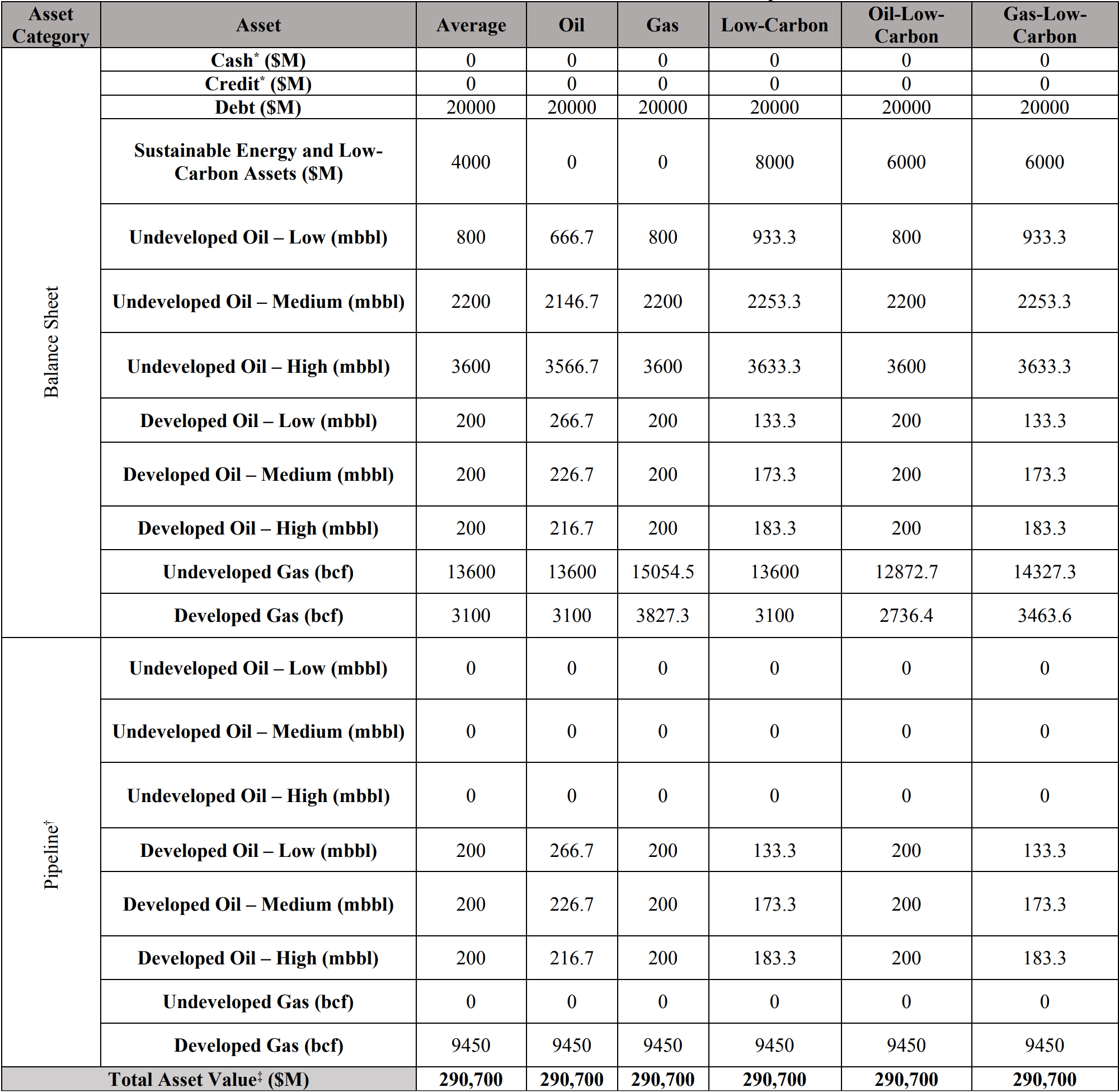}
\caption{IOC action space categorized by the game stage in which they are applicable (cont.). }
\label{tbl:ioc_balance_sheet_distribution}
\end{table*}

\pagebreak

\section*{Appendix A.8 - IOC decision-making metrics}

\begin{table*}[!htb]
\centering
\includegraphics[width=0.82485\linewidth]{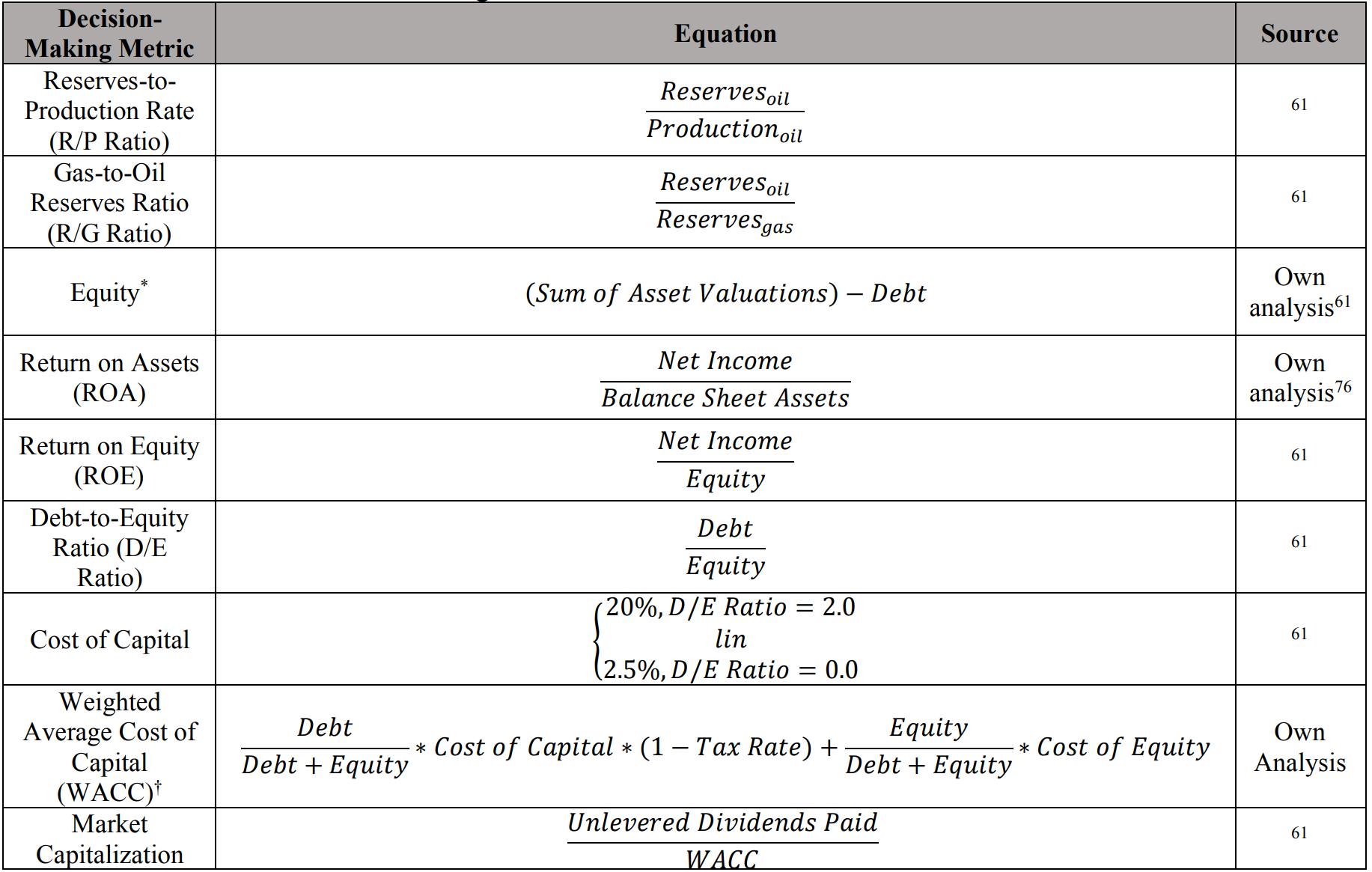}
\caption{$^*$ The costs of oil and gas assets include costs associated with the total amount of developed and undeveloped reserves for an agent in a given year. These costs are only added to an agent's Equity if the oil price remains above the asset's lifting costs (e.g. developed high asset costs are included if that year's oil price is at least $\$40$/bbl); the costs of low-carbon assets are simply the number of low-carbon assets its maintains in respective year, not the cost at which it paid for them via the low-carbon auction or through the player-to-player trading desk. 
$^\dagger$ A global, average corporate tax rate ($24\%$) is used; cost of equity calculated using CAPM method}
\label{tbl:ioc_decision_making_metrics}
\end{table*}


\pagebreak

\section*{Appendix B.1 - Additional selected matches to Figure \ref{fig:average_dividend_payout}}


\begin{figure*}[!htb]
\centering
\includegraphics[width=.9\linewidth]{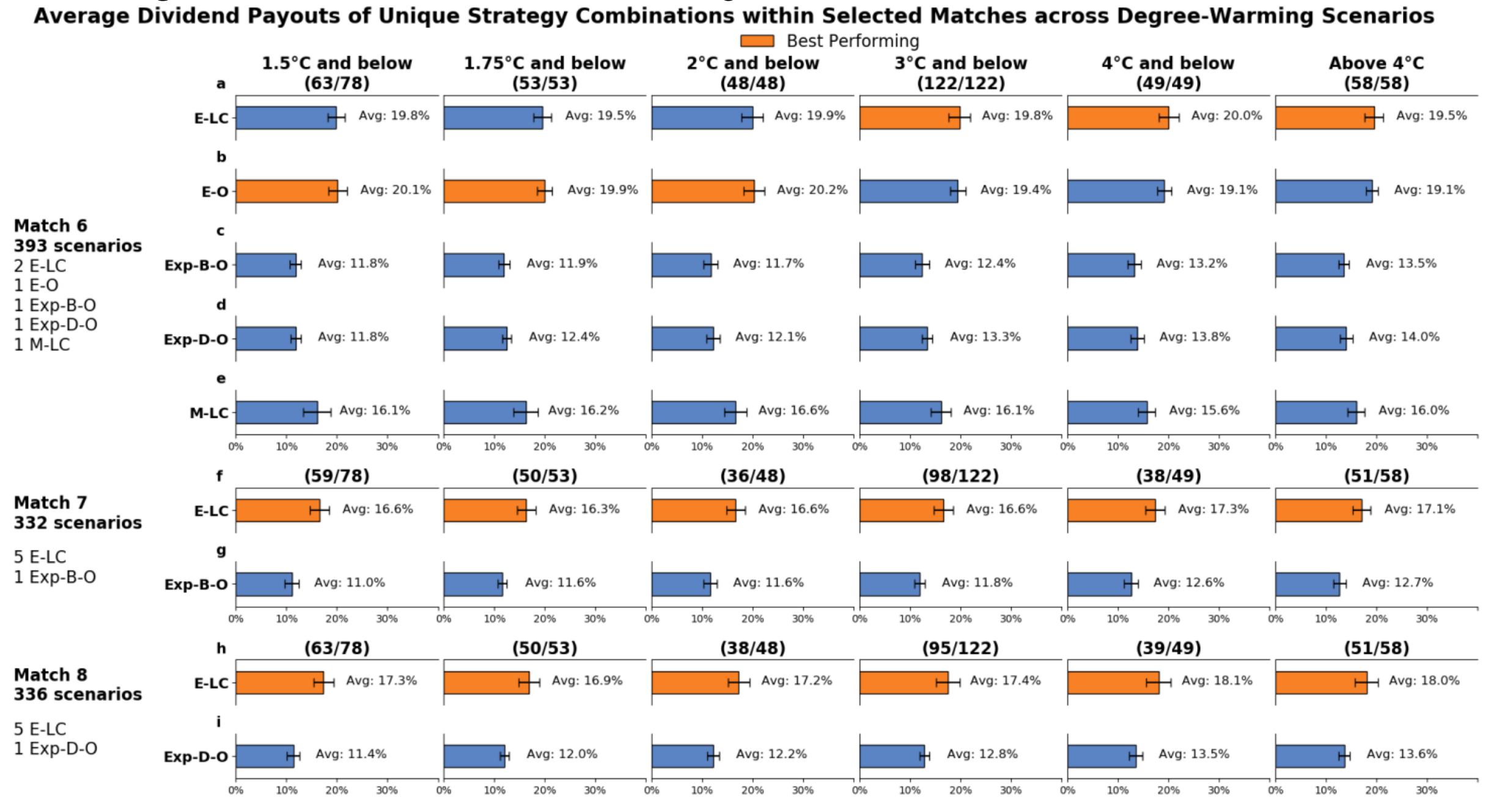}
\caption{\textbf{Average dividend payouts of initial low-carbon movement (Figure \ref{fig:emergent_strategies}d) and business model
(Figure \ref{fig:emergent_strategies}g) strategy combinations within selected matches of similar or differing strategy combinations across degree-warming scenarios.} \textbf{a-e}, Average dividend payouts as compared to total dividend payouts seen in Match $6$ for the average \textbf{(a)} E-LC, \textbf{(b)} E-O, \textbf{(c)} Exploiter-B-O, \textbf{(d)} Exploiter-D-O, \textbf{(e)} M-LC strategy combinations. \textbf{f-g}, Average dividend payouts as compared to total dividend payouts seen in Match $7$ for the average \textbf{(f)} E-LC, \textbf{(g)} Exploiter-B-O strategy combinations. \textbf{h-i}, Average dividend payouts as compared to total dividend payouts seen in
Match $8$ for the average \textbf{(h)} E-LC, \textbf{(i)} Exploiter-D-O strategy combinations.}
\label{fig:average_dividend_payouts_appendix}
\end{figure*}

\pagebreak

\section*{Additional selected matches to Figure \ref{fig:emergent_strategy_combinations}}

\begin{figure*}[!htb]
\centering
\includegraphics[width=.9\linewidth]{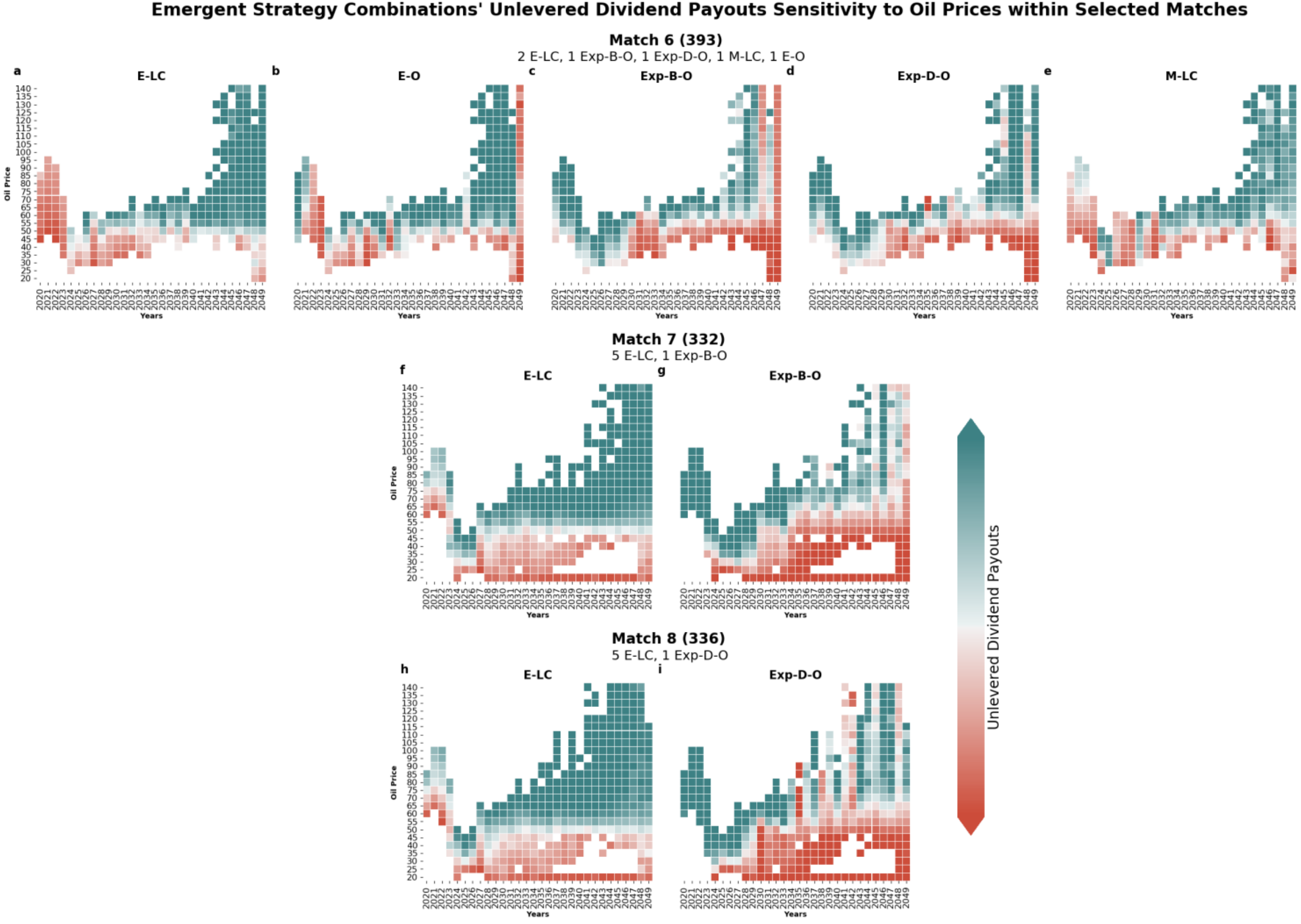}
\caption{\textbf{The sensitivity of yearly dividend payouts to varying oil prices for strategy combinations within
five unique matches.} \textbf{a-e}, Sensitivity of yearly dividend payouts to varying oil prices seen in Match $6$ for average \textbf{(a)} E-LC, \textbf{(b)} E-O, \textbf{(c)} Exploiter-B-O, \textbf{(d)} Exploiter-D-O, \textbf{(e)} M-LC strategy combinations. \textbf{f-g}, Sensitivity of yearly dividend payouts to varying oil prices seen in Match $7$ for average \textbf{(f)} E-LC, \textbf{(g)} Exploiter-B-O strategy combinations. \textbf{h-k}, Sensitivity of yearly dividend payouts to varying oil prices seen in Match 8 for average \textbf{(h)} ELC, \textbf{(i)} Exploiter-D-O strategy combinations.}
\label{fig:emergent_strategies_combinations_appendix}
\end{figure*}




\end{document}